\documentclass{article}

\usepackage[utf8]{inputenc}
\usepackage[british]{babel}
\usepackage{CJKutf8}

\usepackage{microtype}

\usepackage{tikz}
\usepackage{graphicx}
\usepackage{environ}
\usepackage{pdfpages}
\NewEnviron{scale}[1]{%
  \begin{tikzpicture}
    \node[scale=#1]{%
      \BODY
    };
  \end{tikzpicture}
}

\usepackage[hmarginratio=1:1,top=32mm,columnsep=20pt]{geometry}
\usepackage{multicol}
\usepackage[hang, small,labelfont=bf,up,textfont=it,up]{caption}
\usepackage{booktabs}
\usepackage{float}
\usepackage[colorlinks,citecolor=blue]{hyperref}
\usepackage{listings}

\usepackage{abstract}

\usepackage{titlesec}
\titleformat{\section}[block]{\large\scshape\centering}{\thesection.}{1em}{}
\titleformat{\subsection}[block]{\large}{\thesubsection.}{1em}{}

\title{\vspace{-15mm}\fontsize{16pt}{10pt}\selectfont
\textbf{Enabling Learning by Teaching: Intuitive Composing of E-Learning 
Modules}}

\author{
\begin{tabular}{c c c}
    Alexander Berntsen & Stian Ellingsen & Emil Henry Flakk \\
    \href{mailto:alexander@plaimi.net}{alexander@plaimi.net} &
    \href{mailto:stian@plaimi.net}{stian@plaimi.net} &
    \href{mailto:emil@plaimi.net}{emil@plaimi.net}
\end{tabular}
}

\graphicspath{{fig/}}

\begin{document}
\maketitle
\begin{abstract}
\noindent In an effort to foster learning by teaching, we propose the development of a 
canvas system that makes composing e-learning modules intuitive. We try to 
empower and liberate non-technical module users by lowering the bar for 
turning them into module authors, a bar previously set far too high. In turn, 
this stimulates learning through teaching. By making a damn fine piece of 
software, we furthermore make module authoring more pleasant for experienced 
authors as well. We propose a system that initially enables users to easily 
compose H5P modules. These modules are successively easy to share and modify. 
Through gamification we encourage authors to share their work, and to improve 
the works of others.

\end{abstract}
\tableofcontents
\listoffigures
\begin{multicols}{2}
\section{Introduction}
The "Bottom Of the Pyramid" (BOP) is a term central to the development of
value-chains in emerging markets in developing countries. It is understood as
the largest, although poorest segment of the population. The key to economic 
development lies in activating this segment of the population, enabling them 
not only to use modern (digital) services, but to author them as 
well\cite{prahalad2009fortune}.

The inspiration for our idea comes from applying this general framework to the
field of education software. The economics of educational services are 
unfavourable to those who need it the most.

We improve the status quo by offering a novel system for composing e-learning 
modules that is easy to use by non-technical users, but expressive enough to 
benefit power users as well.

This improvement is a stepping stone towards turning e-learning authoring into 
an activity achievable by even school children. It is also a stepping stone 
towards stimulating learning by teaching, a stimuli school children find 
themselves largely deprived of in most situated learning environments, i.e.\ 
schools.

\section{The problem}
Alas, most software is written for a small subset of the population. 
H5P\footnote{\url{http://h5p.org/}} is one of many novel examples of software 
providing added-value to tech-savvy teachers and educational software 
developers. However, the BOP would in this case be the students.

By providing the correct tooling and educational framework, we could enable 
students to act not only as users of off-the-shelf educational software, but 
also as authors of their own or their co-students' learning experience. By
doing so, we could allow the student to feel some degree of pride in the
coursework. Moreover, it would allow for reinforced learning, as well as viral 
effects (sharing), a major force in the modern network economy.

Such a framework would also encourage the students to make services and tools
more adapted to their particular learning situation, improving the overall
user experience. It would additionally be a valuable source of inspiration for 
educational software developers and user experience designers. And if 
successful, usage data would be interesting for analytics research.

Sadly, such a framework doesn't exist. So we want to move H5P closer to being 
one. In order to achieve this we provide intuitive composition of H5P modules, 
and, ultimately, of all kinds of e-learning modules, including new ones.

\section{Our idea}
We suggest developing an engaging ``canvas''  for authoring e-learning 
modules, taking cues from 
\begin{CJK}{UTF8}{min}RPGツクール\end{CJK}\footnote{\url{http://www.rpgmakerweb.com/}}, 
Game Maker\footnote{\url{https://www.yoyogames.com/studio}}, and other similar 
software that succeed in take advantage of an intuitive interface with a 
capacious featureset.

By ``canvas'' (or ``scene graph''), we mean an easy-to-learn system of widgets 
and composite stores (with rich content), and flows connecting them together 
to a cohesive system. The widgets would initially be H5P modules.

\begin{figure}[H]
    \centering
    \begin{scale}{0.1}
        \includegraphics{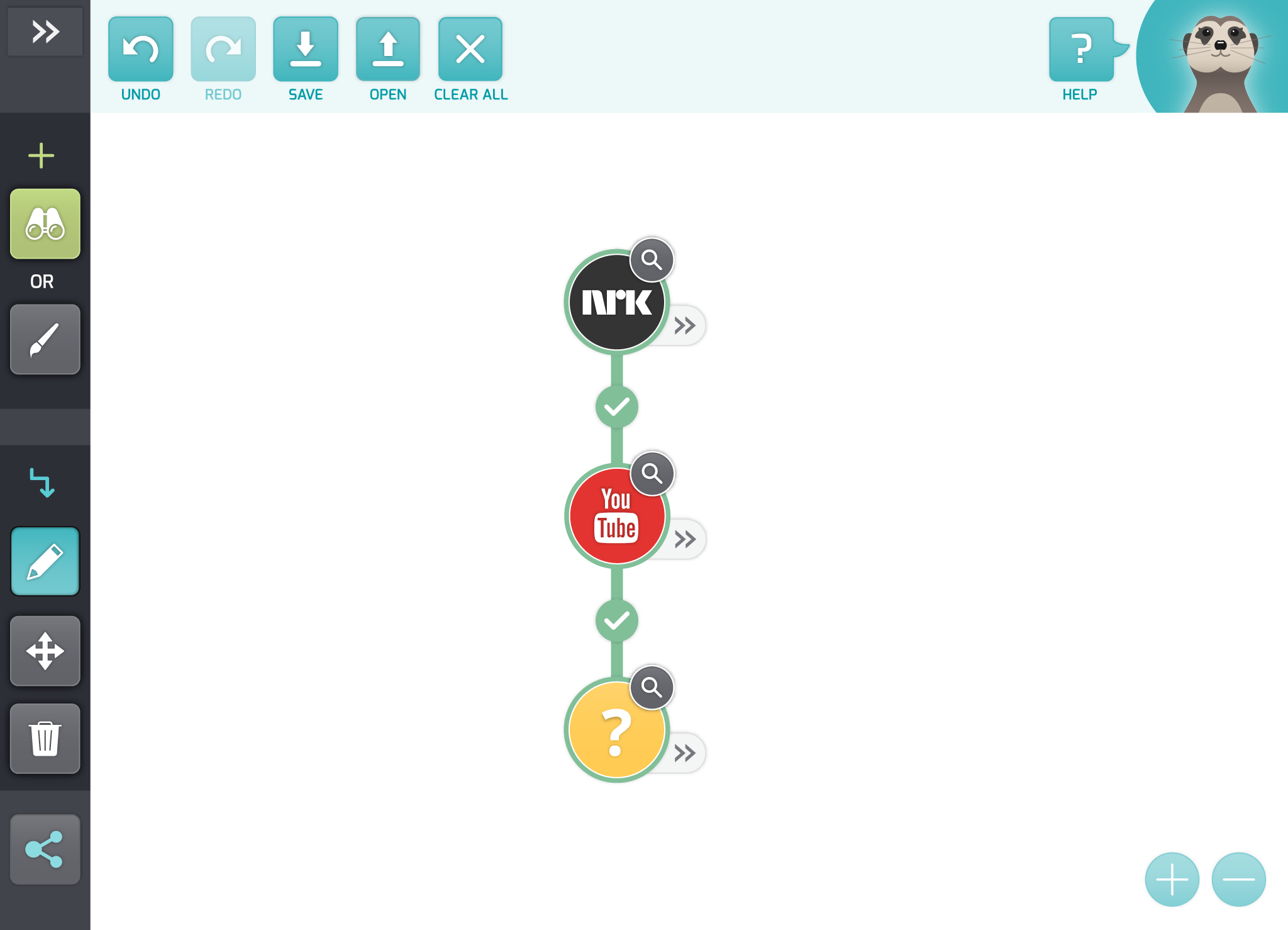}
    \end{scale}
    \caption{A system where the user watches a YouTube video, then reads an 
   article on NRK, and finally does a quiz}
\label{fig:younrkquiz}
\end{figure}

An example of such a system might be a test leveraging media like GNU 
Mediagoblin, Youtube, NRK, Twitter, and Nasjonal Digital 
Læringsarena\footnote{\url{http://ndla.no/}} to showcase the issue at hand by 
embedding video and articles, etc., before testing the user's comprehension at 
the end with an H5P quiz. Such a system is shown in 
Figure~\ref{fig:younrkquiz}. Thus, the student-come-software-developer is able 
to prove their comprehension of multimedia (as mandated by the curriculum), 
and other students are permitted to take part in the learning experience.
Incentives for doing so might be given by the teacher, or by arranging 
collaborations that invite the students to share their best ideas.

Taking cue from Minecraft, we also propose another viable incentive: The 
student is provided with ready-made canvases with actors and stores made by 
other users, and encouraged to make the application perform certain actions 
(``gamification''). This is akin to the redstone system found in Minecraft, 
where even young users are able to construct discrete logic circuits providing 
useful functionality like opening doors, or switching on 
lights\cite{brand2013crafting}.

\begin{figure}[H]
    \centering
    \begin{scale}{0.1}
        \includegraphics{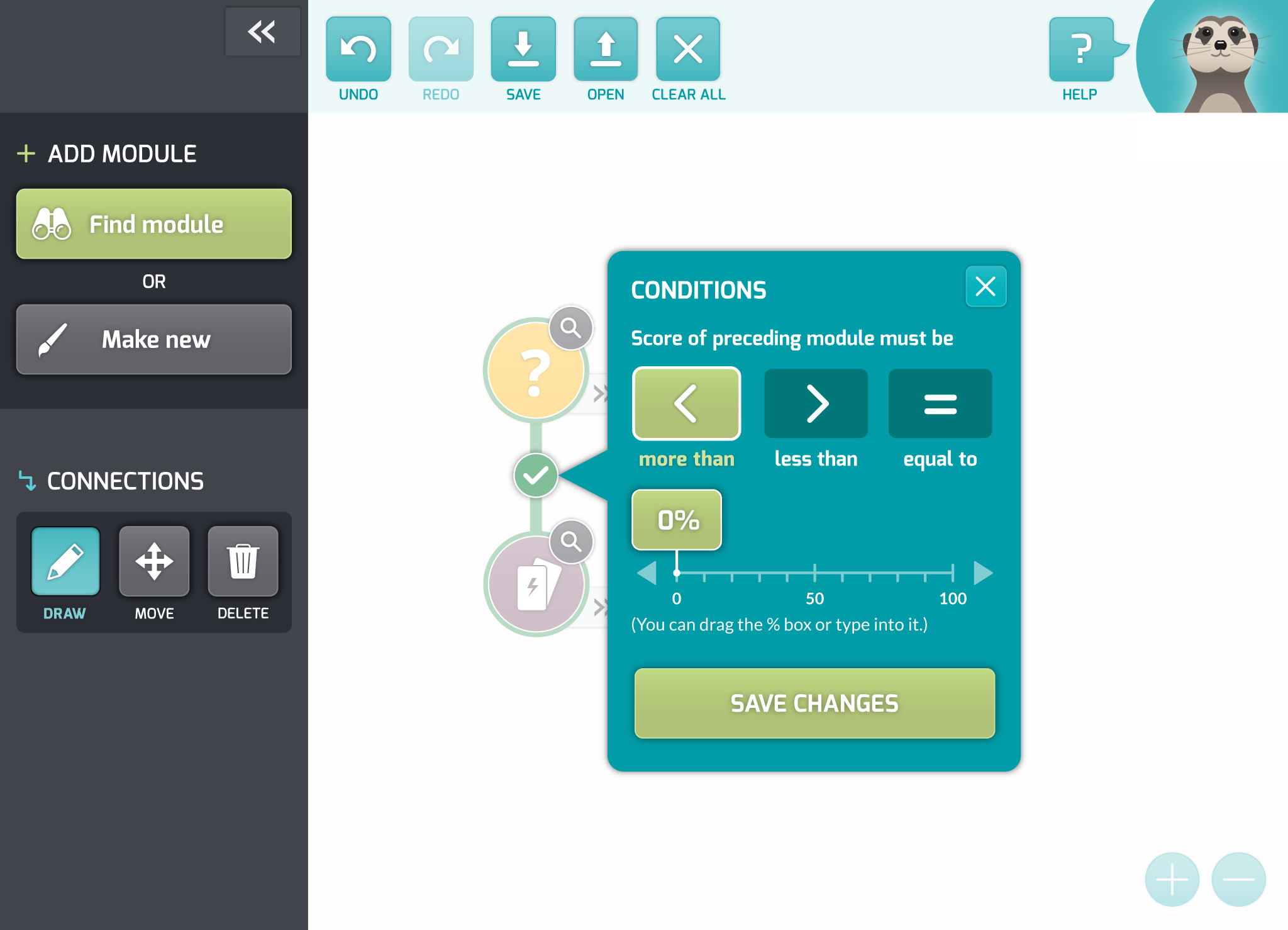}
    \end{scale}
    \caption{A dialogue for tweaking conditional flow}
\label{fig:conditions}
\end{figure}

Consequently, we have customisable flows between widgets. They may be 
customised initially via simple if-expressions on the form $if$ $\chi$ $then$ 
$\alpha$, $else$ $\beta$, where e.g. $\chi=$ 80\% completion rate, and 
$\alpha=$ an H5P memory game, and $\beta=$ an H5P quiz. An example of the 
conditions dialogue is shown in Figure~\ref{fig:conditions}.

Through use of our canvas, we eliminate the BOP by liberating and empowering 
it to make its own user experience. As an added bonus, our canvas may spark 
some latent creative souls, or inspire technological awareness and interest. 
In our increasingly computerised society, this is in itself a noble cause. 

Thus, this canvas might prove to be a force for bridging the gap between just 
being a computer user, and having a promising future career in computer 
software. While Computing At 
School\footnote{\url{http://www.computingatschool.org.uk/}} have had great 
success in the UK, there is as of today no readily-available path for 
acquiring the advanced knowledge needed to develop modern systems given the 
current education system in Norway. Grassroots organisations such as Lær Kidsa 
Koding\footnote{\url{http://www.kidsakoder.no/}} are doing good work, but have 
yet to strongly influence the education system. If our canvas is picked up by 
prominent e-learning providers like Nasjonal digital læringsarena we can 
liberate and empower users through direct action, circumventing bureaucracy.

In addition to users teaching themselves technology, they also teach 
themselves the curriculum more effectively. The student becomes the teacher, 
and we achieve learning by teaching, an often sought-after method of 
reinforced learning. By only being e-learning module \emph{users}, students 
are limited to learning through observation, experimentation, and (to some 
extent) mistakes. Learning through teaching offers advantages not possible to 
fully realise through either of these; advantages that won't manifest if the 
student is exclusively relying on an external 
teacher\cite{cortese2005learning}.

Our canvas makes composing e-learning modules intuitive for non-technical 
users and at the same time powerful enough to entice power users and 
established e-learning module authors. Consequently, the target demographic of 
our canvas is not limited to the BOP, but extends to include current 
e-learning module authors.

\begin{figure}[H]
    \centering
    \begin{scale}{0.1}
        \includegraphics{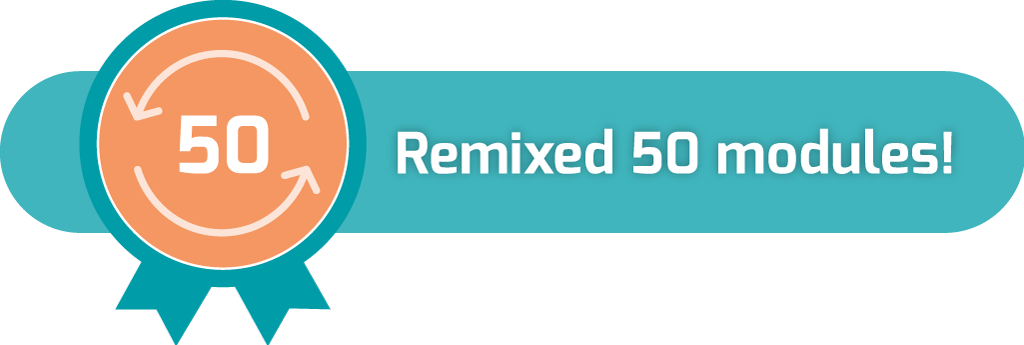}
    \end{scale}
    \caption{A reward for positive behaviour}
\label{fig:rewards}
\end{figure}

To further underline our empowering of the BOP, we make sharing of works very 
simple, and make it equally simple to author derivatives (forks) and 
meta-works (collections). Incentive for sharing your work, and improving or 
remixing the work of others is provided through gamification of the canvas. As 
an example, there may be a reward system for users whose modules are often 
remixed, and for users who make an improvement to a module that then gets 
integrated back into the original module itself. These rewards are then 
presented in a positive and motivating way to the user, as shown in 
Figure~\ref{fig:rewards}.

With gamification, we can also help ensure high quality modules. With a rating 
system and achievements for obtaining a high rating, we realise a 
self-regulating community.

Initially we aim to support authoring H5P modules, focussing our development 
at H5P integration. But with a good modular design we can extend our canvas to 
support other standards as well in the future. We also want to support 
embedding of things like videos and news articles. Lastly, careful attention 
must be paid to the application programming interface, so that it is easy for 
future e-learning modules to support our system.

\section{The details}
\subsection{Concept }
\begin{figure}[H]
    \centering
    \begin{scale}{0.1}
        \includegraphics{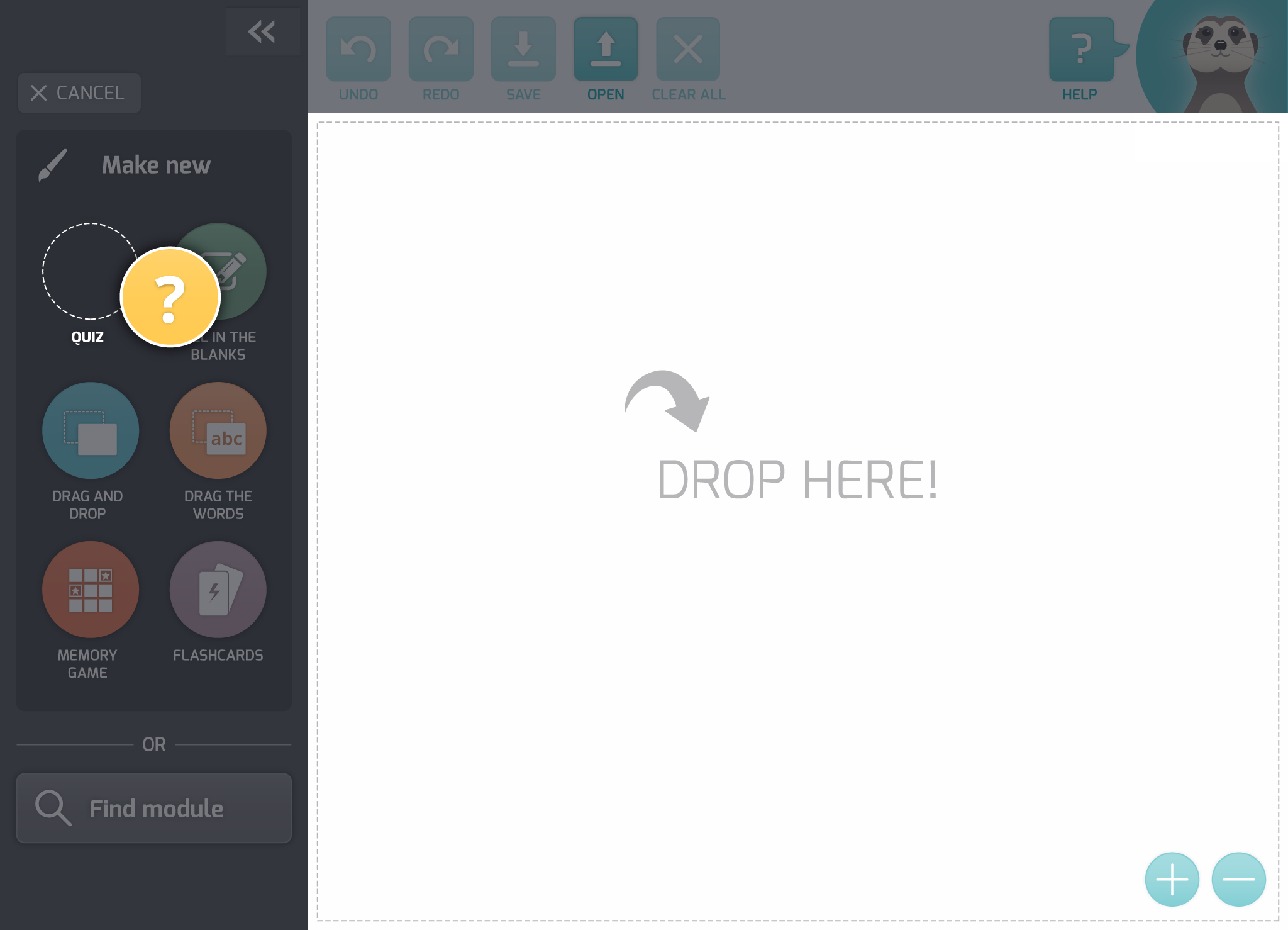}
    \end{scale}
    \caption{A user dragging a module onto the canvas to add it}
   \label{fig:adding}
\end{figure}

We propose a canvas system wherein the user may drag e-learning modules from a 
toolbox, and drop them onto the canvas, as in Figure~\ref{fig:adding}. The 
modules may be empty or finished. If they are empty, then the authoring tool 
for said module opens. If it is not empty, then the user may further edit it 
if they wish. In addition to drag-and-drop, the user may also use an ``add'' 
button, as drag-and-drop is not viable for universal design.

\begin{figure}[H]
    \centering
    \begin{scale}{0.1}
        \includegraphics{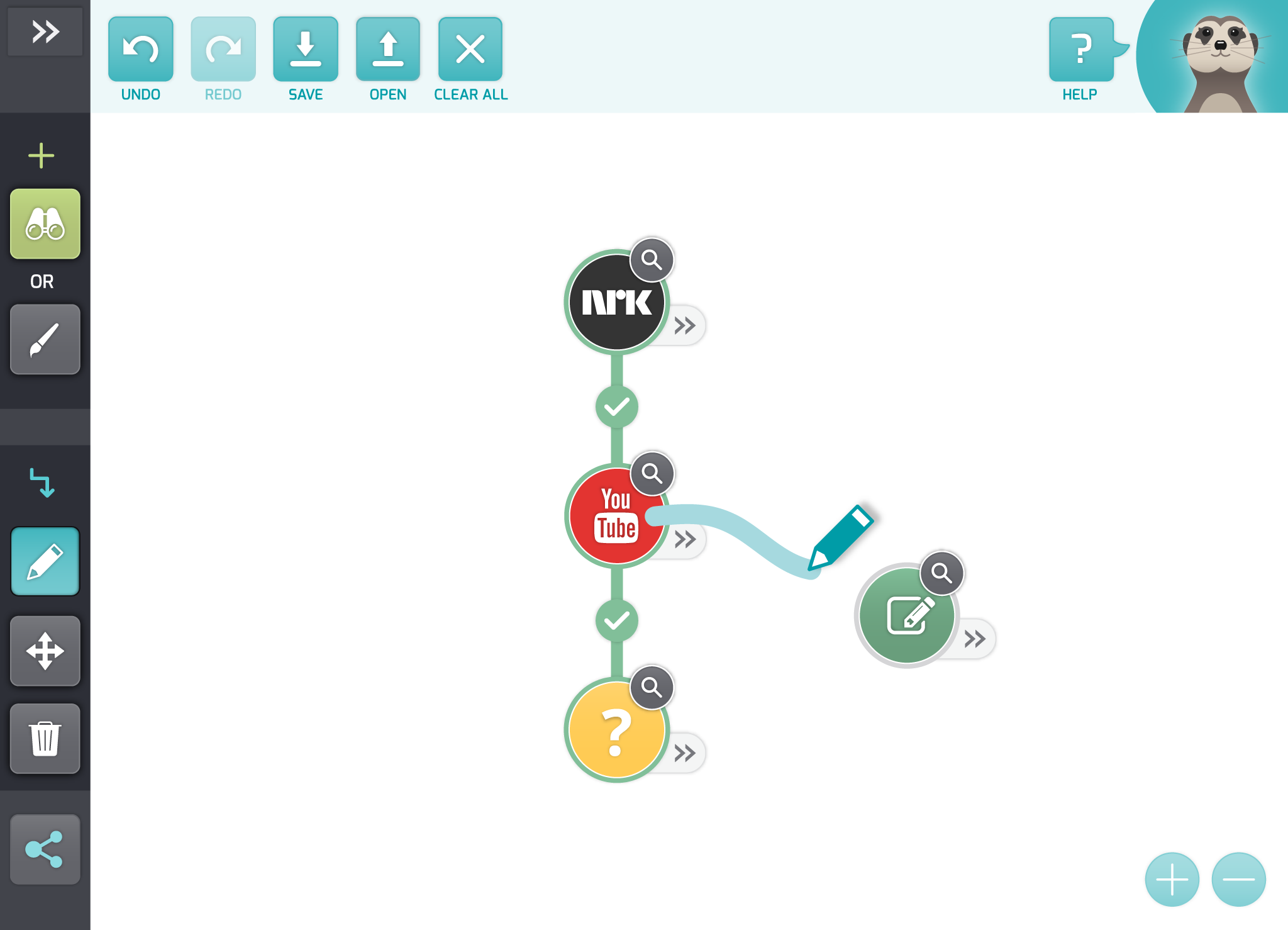}
    \end{scale}
    \caption{Visualised control flow}
   \label{fig:arrows}
\end{figure}

To schedule data and control flow, the user drags arrows between the modules 
on the canvas. These associations may be customised. ``If the user finished 
[module a] with a score of more than 80\% correct, then direct them to [module 
b]; if the user did not, then direct them to [module c]''. A simple 
illustration of this is in Figure~\ref{fig:arrows}.

Compositions are modules themselves.

\begin{figure}[H]
    \centering
    \begin{scale}{0.1}
        \includegraphics{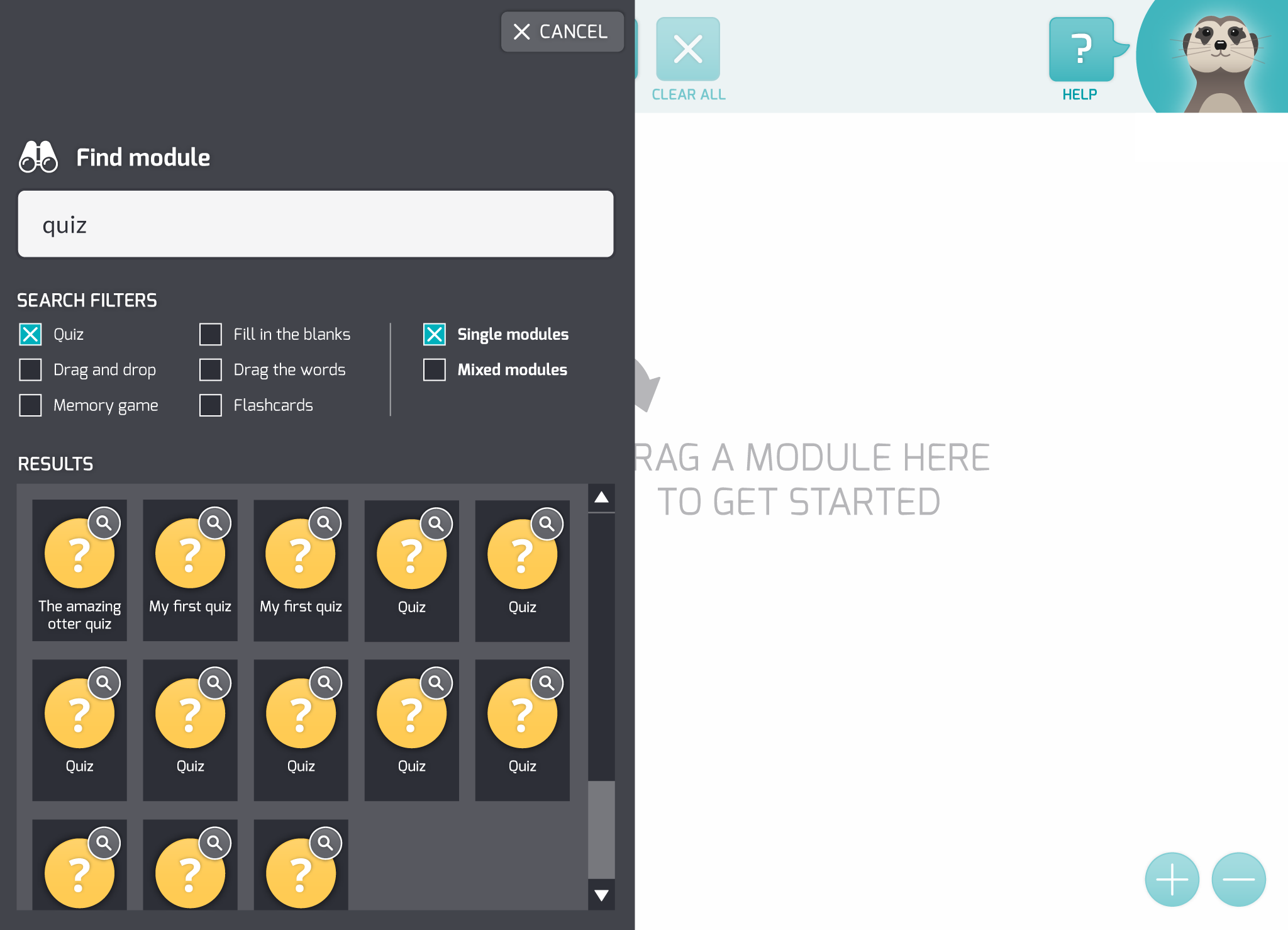}
    \end{scale}
    \caption{Search results, filtered by type}
   \label{fig:search}
\end{figure}

When a user wants to compose modules, users may search for which modules to 
use. We want to encourage reuse and remixing rather than novelties. 
Consequently we focus the search option before the ``author new module'' 
option. The search option is shown in Figure~\ref{fig:search}, which also 
shows results filtering in action.

As another incentive, when a user authors a new module, we track reuses and 
remixes. Users may in a presentation of all their authored modules see how 
many reuses and remixes have occurred, and follow links to these. If they like 
a remix they may merge the changes between the remix and the original.

Compositions may be derived in whole, and further developed by another author. 
Users may merge modifications to these as well. Consider an English Language 
test for a school curriculum. A teacher at a different school (or the same 
school next year) may want to reuse this test since they have the same 
curriculum. But they might furthermore want to modify it. If the original 
author finds these modifications useful, then it may merge them to the 
original.

\begin{figure}[H]
    \centering
    \begin{scale}{0.1}
        \includegraphics{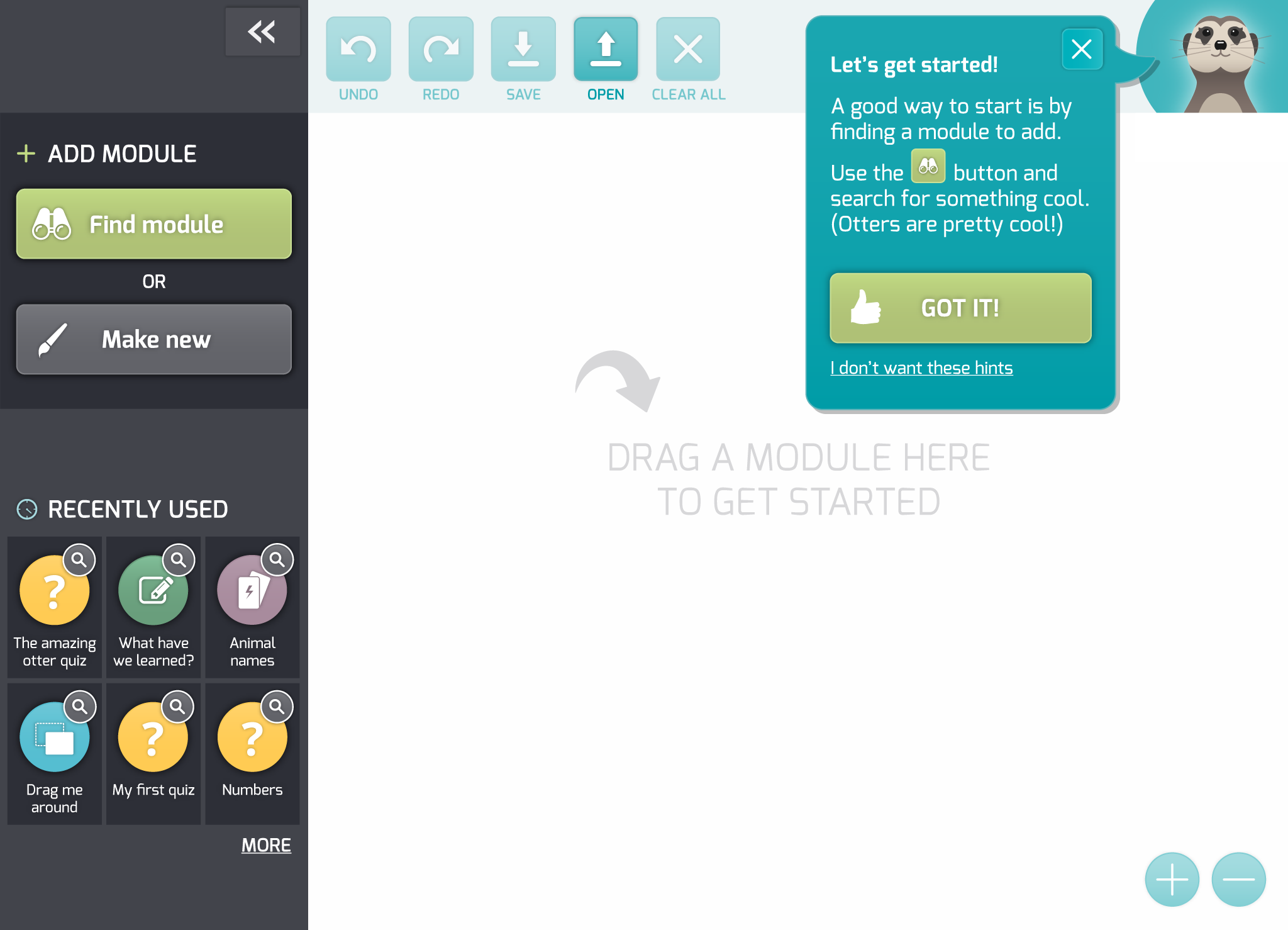}
    \end{scale}
    \caption{An avatar helping the user}
   \label{fig:avatar}
\end{figure}

Avatars have been successfully used before to motivate 
children\cite{gossen2012search}. We propose the use of an avatar for 
explaining the interface and pointing out problems with the compositions, such 
as integrity issues, e.g.\ ``the composition never ends'', or continuity 
warnings, e.g.\ ``this module is visited two times'',. The avatar's appearance 
may be customised. We show an example of an avatar dialogue in 
Figure~\ref{fig:avatar}, although the avatar itself is merely a placeholder.

For our actual avatars, we have chosen otters. plaimi's mascot is an otter, so 
this decision is done partly for cohesive branding concerns, but primarily for 
the same reasons it was chosen as plaimi's mascot in the first place: the 
otter is a noted playful and social creature\cite{gordon1908otter}. It is 
furthermore considered cute by the general population, but capable of being 
ferocious\cite{belanger2011review}, which is appealing for the target 
demographic as girls tend to prefer cute avatars, whilst boys prefer rude 
avatars\cite{inal2006children}.

To encourage certain behaviour, the avatar may reward positive behaviour with 
a badge, e.g.\ ``200 remixes!'', or ``remixed 200 modules!''.

Punishing bad behaviour rarely works, and punishment is easily evaded. Instead 
we simply prohibit bad behaviour. Chatting is done via the avatars. To 
communicate with another user, you visit their profile and click on their 
avatar to bring up a dial menu for constructing a query like ``you should add 
[module] to your composition!'', or ``I like this module!''. This makes 
localisation much easier since these set sentences may be translated 
accurately. It also makes our system safe for 
children\cite{sadler2012virtual}. This is a recent trend in computer games 
such as Journey and Hearthstone for the same reasons. This eliminates hate 
speech\cite{hearthstone}, which is not acceptable with a child audience.

Users may, via their avatars as just described, encourage other users by 
telling them that they like their modules. There is no way to dislike a 
module, as that might be demotivating for children. Avatars as well as modules 
may be favoured. The user's favourite modules and avatars are more easily 
accessible than the rest.

Users sign up with a log-on ID and password, and optionally an email address 
for recovering forgotten passwords. The users name their avatars which is then 
their screen names. The avatar has a name. There may be duplicate names. 
People have duplicate names in real life too [citation needed], and society 
hasn't collapsed yet. When choosing a name, feedback is given if the name is 
not unique.

\subsection{Requirements}
We place high usability requirements on our canvas system in order to enable
school children to use it.

Our target demographic at its largest is pretty much everyone that wants to 
learn or teach something. Therefore universal design is of the utmost 
importance. It is furthermore legally required in Norway\cite{uuforskrift}. 
More importantly it is The Right Thing to do. Universal design is difficult 
with such a graphical system, but of the utmost importance. Compromises shall 
err on the side of caution and favour universal design to expressiveness.

Dealing with children has interface design implications. Tap actions should 
allow some slack with bigger hit boxes than actual buttons and similar 
measurements. Splash screens are bad, mkay. Turns out kids don't have that 
great an attention-span. So get to the point quick as. These things are 
discussed in more detail in Section~\ref{principles}.

In order to encourage reuse and remixing, modules and assets are distributed 
under a free licence, CC-BY-SA\@. This also sets a positive example for school 
children to become good citizens of our society. Indeed so does the entire 
project by being licensed as AGPL\@\cite{educational}.

\subsection{Functional Details}
\subsubsection{Frontend}
For our canvas system and frontend in general we primarily use the Elm 
programming language\cite{czaplicki2012elm}.

Elm is an immature language with several known 
shortcomings\cite{smitsextreme}, but the more popular alternative JavaScript 
is a mature language with even more known 
shortcomings\cite{flanagan2006javascript}. Although early research suggests 
that using Elm will not magically result in a better 
product\cite{buist2014extending}, it does allow for purely functional 
programming of denotational graphical interfaces\cite{czaplicki2012elm}, which 
has several known advantages for interactive graphical 
programs\cite{berntsen2014quest}. Elm itself has several specific advantages 
as well --- and excellent tooling to boot\cite{kraeutmann2015functional}.

Being an immature language, not all libraries we could possibly wish for are 
readily available. Therefore accessing JavaScript directly is still useful at 
times. Thankfully, Elm comes with a sophisticated JavaScript foreign 
functional interface\cite{elmports}. And if it becomes necessary to use 
JavaScript libraries to interact with JavaScript primitives presently outside 
of Elm's reach, we may fall-back on JavaScript for this.

\subsubsection{Backend}
For dealing with all things backend we use the Haskell programming 
language\cite{marlow2010haskell}. Haskell is a mature and widely used 
programming language with advantages similar to that of Elm, except more 
advantages and more powerful advantages. Interesting advantages over Elm 
include laziness for modularisation\cite{hughes1989functional}, typeclasses 
for elegant ad-hoc polymorphism\cite{wadler1989make}, and an in general more 
powerful and interesting type system\cite{jones2003wearing}. Lastly, Haskell 
seriously equips you for all cost-models of parallelism and 
concurrency\cite{jones2012future}.

We write property laws and test these using 
QuickCheck\cite{claessen2011quickcheck}, a well-established tool used in 
industry\cite{arts2006testing}.

Haskell, being a mature language, provides us with several useful libraries to 
ensure a shorter time to market.

\subsubsection{Reuse and remix}
We want to encourage reusing and remixing. We define reuse to be using someone 
else's module as-is, and remix to be using someone else's module with 
modifications.

Care must be taken not to merely give incentives for having your modules 
reused and remixed by others, as this fosters a culture where being the author 
of novelties is the most rewarding. Rewards should therefore be doled out more 
handsomely for actually reusing and remixing than for authoring modules that 
are successfully subsequently reused and remixed. This also makes sense since 
the latter is accumulatively (having your module be reused and remixed is 
tracked and awarded recursively) rewarding passive behaviour, whilst the 
former rewards active behaviour. 

\subsection{Design document}
Given our emphasis on usability, we have had a design document authored by a 
user experience developer. This document includes a graphical profile, some 
specific user-interface design, and some overarching design principles. It may 
be found in its entirety in Appendix~\ref{design}.

\section{First principles}
\label{principles}
We wish to discuss \textit{learning} itself, and how this knowledge affects 
our system. A common technique used in most of these systems is \textbf{spaced 
repetition} combined with \textbf{testing}, i.e.\ the pacing of repetition 
throughout time. This approach is well-proven by research, and could be 
implemented both as a part of our services (allowing module authors to apply 
it to their own students), or as a part of the experience of learning the 
tools. An example might be: `After 5 days, please repeat this factoid and test 
the student.` Of course, such an approach might prove cumbersome, and 
techniques like machine learning might automate it in some meaningful fashion 
using a data-oriented approached. Reminders are most effective when occurring 
just before the memory fades. Successive reminders will consequently be 
separated at longer and longer intervals\cite{memrise}.

In testing, there exists mainly three techniques for assessment:

\begin{itemize}
\item Reading, which is the reproduction of textual representations of the fact.
\item Multiple choice, where one is allowed multiple alternatives, where at
least one of them should be true. Often one or more of the alternatives are `the
odd one out', to test the student's ability to discriminate irrelevant 
concepts.
\item Generation, where the user is made to produce something, for example
filling in a word. This technique is highly efficient with regards to 
retention, especially when combined with quick 
feedback\cite{potts2014benefit}.
\end{itemize}

Thus, we consider generation. An example of this might be generating a formally
valid blueprint of some e-module, and asking the user what it does. Or it might
be to make the user connect the relevant parts to make it work as desired. Thus,
we foster true comprehension of the system mechanics and avoid cramming. As
noted in the literature in the Memrise review, errorful generation might 
actually be beneficial, and the user should not be punished for making 
mistakes. Rather, we suggest to embrace it and make failure as smooth as 
possible, with quick iterations.

A big part of any learning system is the \textit{metaphor}, or the user
interface. The system needs to provide some incentive for the user to perform
some action. Thus, the user interface will always be a question of demographics.
We suggest that the user interface should be amenable to at least some
simplifications to scale down to the technology-native elementary school
demographic. Some concrete suggestions are:

\begin{itemize}
\item The user interface needs to take into consideration the rather poor motor
skills of young children\cite{kidsquora}. Thus, tap actions should not be that
precise or responsive, but instead allow for some slack. For example, multiple
taps should not result in multiple windows, but rather provide a time buffer for
the action to launch. Also, the hit box of buttons should be greater than their
actual size.
\item Avoid splash screens\cite{kidsluke}, and in general get to the point as 
quickly as
possible. Thus, it might be better to provide a simpler, but usable interface
rather than a complete experience first-hand. E.g.\ one may wish to avoid 
modal dialogues, as they provide children which textual choice, which they
might not understand or be interested in.
\item To provide some sense of continuity, and also a way of accumulating
rewards, the user constructs an avatar. This concept is known from video
games, and has already been suggested to be a viable way of familiarising
children with complicated services like search engines\cite{gossen2012search}.
\end{itemize}

Unfortunately, we were not able to locate high-quality sources of research in
this area. Most sources are based on de-facto and ad-hoc solutions, with little
or no scientific basis outside of in-house usability studies. Thus, we are
potentially missing out on a great deal of useful data. On the other hand, 
this means we have the unique opportunity to not only be innovators, but also 
inspiring flagships.

\textbf{Takeaway points}, based on previous discussion and the reviews in 
Section~\ref{related}.

\begin{itemize}
\item Spaced learning with testing using generation fosters comprehension and
recall. We need to embrace generating errors as a way of learning.
\item Quick and simple feedback with additive gains facilitate user engagement
and learning greatly, and promotes attention continuity.
\item The system should provide a simple (preferably physical) metaphor as part
of its user story.
\item The system needs to use open, free standards to enhance composition and
foster a culture of reuse and experimentation.
\item We should strive for feature parity between different clients, to simplify
the user story and provide a consistent experience.
\item A strong, data-driven approach involving the users strengthens community
relations and improved quality by providing relevant feedback to create,
enhance, or QA existing material.
\end{itemize}

So the big question then becomes what do we do with all these takeaway points. 
In order to provide a cohesive experience for authors as well as users we need 
to take all of this into account. But cohesion traditionally proves difficult 
when coupled with agnosticism --- i.e.\ we would like to compose many --- 
different --- kinds of modules, and different kinds of modules likely have 
different approaches to authoring.

What we are eventually getting at is the need for high quality control in what 
modules we allow, and likely a fair bit of customisation of the ones we do end 
up allowing. Alternatively new module authoring software may be developed 
altogether.

\section{Related work}
\label{related}
H5P has rudimentary editing support. It is unfortunately low-level. But at the 
same time it appears to be easy enough to integrate, which is ideal for our 
system. At any rate, H5P does not offer high-level composing of modules, which 
is our real novelty.

There does not appear to be any mature, publicly available e-learning module 
authors that achieves the sophistication we would enjoy seeing, capable of 
making genuinely diverse modules, from flashcards to memory games, from 
quizzes to mini-games. This is another misfortune.

Intuitive composing seems virtually unheard of. And a standalone 
module-agnostic composing tool appears to be a complete novelty.

However, there exist several authoring tools for the specific things we wish 
to accomplish. I.e.\ there are flashcard authoring tools, quiz authoring 
tools, etc. In this section we discuss those, as well as authoring tools for 
completely different things, and unconventional "authoring tools" like 
Minecraft.

\subsection{Evaluation of Anki}

Anki is a system for spaced learning. It works by organising facts (in a wide
sense, like foreign words or parts of a map) into individual cards, which in
part constitute a deck of facts about a given topic. The user plays through the
deck by going through each card and stating how easy it was to recall the fact.
The software then replays cards depending on how good or bad the recall was
until some threshold of recollection has been met. Bad recall would result in 
the card appearing more often, and vice versa. This algorithm is inspired by the
SuperMemo system written in the 1980s\cite{ankimanual}. The intuition is that 
we reinforce learning by recalling something just as we are about to forget 
about it.

Users are allowed to share, amend, or extend existing decks. While Anki
primarily exists for the purpose of individual learning, the composable and
free software format allows for others to take part in the learning 
experience. The software has enjoyed great popularity amongst several groups, 
most notably learners of foreign languages.

Anki is available both as a desktop client and through a web interface --- 
AnkiWeb\footnote{https://ankiweb.net/}. While the latter easily allows for 
online sharing of decks, it doesn't allow for importing them. This is because 
it only synchronises with the desktop client. This highlights the challenges 
faced by desktop and web clients with different capabilities, and the value of 
isomorphic implementations.

AnkiWeb allows for synchronization and sharing of card decks (as files) using 
a free online service. It is also able to play through the decks, but not 
import them from stand-alone files.

The desktop client of Anki will therefore be the main focus of this evaluation,
as it is the only full implementation of the system. Similar software in the
same area are SuperMemo and Fluxcards.

In terms of licensing, Anki is well-aligned with our project. It is freely 
licensed under the terms of the GNU AGPL v3+. Sub-components are licensed 
under a variety of licenses, mostly GNU GPL and flavours of BSD/MIT licences. 
All art assets are also freely licensed.

\subsubsection{Expressiveness}

Anki allows for the authoring of (decks of) flashcards using standard HTML
components with CSS styling. As a result, users are able to enrich cards with
customised formatting, sound, and video. The users are also able to record their
own voice (and repeat sounds) during playthrough. This offers an interactive
experience to the end-user. Since HTML and CSS are relatively flexible and
well-documented technologies, and is used by many other prominent parties, 
this is likely a good choice by the Anki developers, that gives the users a 
good balance between expressiveness and ease of use, as well as familiarity.

\subsubsection{Ease of use}

Because the technologies used by Anki are freely available and readily
documented, they constitute an open standard. This open container format makes
Anki composable, and since the technologies are rather common (at least HTML),
users can make relatively simple changes without any major obstacles.

However, Anki is considered to have a poor user interface interaction story 
for common scenarios\cite{pcworldanki}.

An example of a problematic story is the rather common case of wanting to share
a deck. This usually has to be done through AnkiWeb, a free online service, 
which allows the user to download new decks made by other users. However, no such
suggestion is made to the user. Thus, the social dimension of Anki is 
downplayed.

Similarly, when using the software for the first time, the user is greeted by 
a shallow window with a single deck, and no explanation of Anki or its 
algorithm is given. The user can click a simple built-in test deck to play 
through it, but this is not suggested by the software, and the deck does not 
showcase Anki's capabilities either (for example the use of rich media or 
LaTeX equations).

Other examples include the main window, which has both a menu bar and some
buttons. One of the buttons suggests that the user might be able to import new
decks, but turns out to be about constructing new decks.

The most complex interface in Anki is the library view. It allows for somewhat
non-trivial querying of all the decks in use, with numerous built-in categories
and tags for distinguishing them (be it difficulty, topic, or the decks
scheduled for today). Also, this rich view is not offered in AnkiWeb, so there's
an inherent divide in the capabilities of searching already imported decks and
looking for new ones, limiting the ability to explore.

Universal design isn't guaranteed since the decks are made by users. There 
doesn't appear to be any way of filtering decks based on accessibility, nor 
any user-interface encouragement to \emph{do} universal design when authoring 
decks.

\subsubsection{Takeaway points}

\begin{itemize}
\item Anki offers spaced repetition learning, which is an established and 
  useful way of learning.
\item The simple metaphor (a deck of flashcards) is intuitive and easy to 
    explain to users\cite{burke2002s}, and simple enough to implement.
\item Anki's use of HTML elements provides a universal container format and
  facilitates sharing. However, the web interface is very limited, and while it
  does allow for sharing, it does not allow for importing decks, thus not
  providing a full (isomorphic) user experience across platforms.
\item The desktop client has a problematic user interface story, and does not
  provide enough directions or examples for effective use. Modal dialogues and
  context switching might provide a more elegant solution for capturing common
  usage patterns (compared to displaying the full set of choices and categories
  available to the user, e.g.\ in the library view).
\end{itemize}

\subsection{Evaluation of Memrise}

Memrise\footnote{\url{http://www.memrise.com}} is a flashcard-based learning 
system. The cards make use of (often crowd sourced) mnemonics combined with 
spaced repetition (akin to Duolingo and Anki) to augment the learning 
experience. It is unique in part due to its openness regarding its techniques 
and its close ties with academic researchers, who are allowed to analysis its 
datasets.

Memrise itself suggests three basic techniques:\cite{memrise}

\begin{itemize}

\item \textbf{Elaborate encoding.} I.e., the construction of cognitive 
  structures of already known material that new knowledge might "attach" 
  itself to. E.g., taxonomies (like the colours red, white, blue) are easier 
  to remember than a random list (astronaut, velvet, cigar). To this end, 
  Memrise uses crowd sourcing (with machine-learning techniques) to suggest 
  "mems", which are the aforementioned cognitive structures to boost learning.

\item \textbf{Choreographed testing}, which test recall and comprehension. Memrise varies
  between simple one-off challenges ("casa = ?") and multiple-choice questions
  to keep things interesting.

\item \textbf{Scheduled reminders}, which is their implementation of spaced 
    learning.

\end{itemize}

Research around — or by — Memrise mostly revolves around the benefits of 
testing, error generation (vis-à-vis inerrant learning), and how spacing 
affects the efficacy of the learning programme.

Testing is the act of challenging retention while studying, which is 
beneficial as part of the learning process. There is no consensus as to why it 
works, but some explanations include that it increases the storage strength of 
memory, and that it generates additional cues that creates (potentially more 
efficient) routes through memory\cite[p.6]{potts2014benefit}.

Generation (with feedback) is an active form of learning, and requires the
student to complete something, for example a sentence ("Lisa is a \_\_\_"),
before getting to know the answer. This is contrasted to reading, where the
student absorbs textual material and definitions, and multiple choice, which
amongst other things tests the ability of the student to see how the different
alternatives relate to each other (finding the odd one out etc.).

Errant learning has been thought to be detrimental to learning as people
recall their errors to a greater extent than their correct
answers\cite{potts2014benefit}, but this research has mostly been done with
already memory-impaired populations, which might have different requirements.
Inerrant learning does not seem to be as effective as error generation: 
``\dots'' generating responses followed by feedback is helpful to memory 
even when many errors are generated, compared with inerrant studying without
generation\cite[p.54]{potts2014benefit}.

Generating errors may benefit vocabulary learning even when there are no 
established associations to the items in 
question\cite[p.54]{potts2014benefit}. While completing sentences appear to 
have the biggest advantage over only reading, other types of generation also 
improves test scores when compared to reading by 
itself\cite[p.73]{benassi2014applying}.

Several experiments have been designed to test assumptions regarding
these models. They are mostly language-orientated (a good fit for a 
programming language!), and might use techniques like archaic words (which are 
certain to be unknown for most test subjects) to provide a "clean slate" for 
testing comprehension and retention of definitions. Research shows that 
participations' perception of learning is influenced by error generation, and 
that participants seemingly perceive generate items as more difficult to 
learn, which may lead to participants putting in more effort to learn the more 
difficult items, which in turn may lead to higher 
retention\cite{potts2014benefit}.

\subsubsection{Expressiveness}

More or less any knowledge that needs to be internalised can fit within the
style of testing offered by Memrise. Users are allowed to create their own
courses, which may contain multimedia levels. For example, users may embed 
videos from Vine\footnote{\url{https://vine.co/}} inside ``mems'' to boost 
retention. 

\subsubsection{Ease of use}

Memrise offers an easy to use interface with a few, mostly obvious buttons. 
Flow doesn't seem to be problematic. Interactive modals guide the user through 
things like creating ``mems'' for their own courses and so on.

\subsubsection{Takeaway points}

\begin{itemize}
\item Generating (that is, having the student do something) is provably more
  efficient than either reading or quizzing.
\item We need to embrace generating errors as a way of learning, and give feedback
  as quickly as possible to promote retention and comprehension.
\end{itemize}

\subsection{Evaluation of Duolingo}

Duolingo\footnote{\url{https://www.duolingo.com}} is a solution that provides 
spaced, paced learning. Users are presented with different challenges, usually 
centered around words (for example their pronunciation), and are able to 
advance through a hierarchy of lessons. Upon completion of lessons, modules, 
levels etc., users are granted experience points, which can be spent buying 
more attempts, in-game apparel, and so on. On completing levels, the overall 
retention of the lesson is noted using a strength bar. As time goes by, this 
retention decreases, and the user is prompted to re-take lessons, not 
dissimilar to the system provided by Anki.

Users are able to perform simple tasks like translating individual words or
sentences, pronouncing them correctly (using a microphone), conjugating nouns
and verbs, and so on. These tasks usually come in sets, and such a set in 
turns constitutes a module or level in the game. However, progression need not 
be linear, and more proficient users can opt to perform tasks qualifying for 
more advanced modules early on if so desired.

When performing a task, the user has a given set of hearts or HP (hit points, 
an analogy from games). When they fail a task, the count is decremented. Users 
can also opt for timed lessons.

\subsubsection{Expressiveness}

Duolingo is proprietary software that does not offer a way for users to author 
their own assets, and instead seeks to keep them content by providing it with 
the program.

Duolingo allows users to translate texts to (im)prove their language
proficiency, and uses community filtering to ascertain the quality of the
translations. This is how the service pays for itself --- as a mechanical 
turk for natural languages, currently partnering with CNN and
Buzzfeed\cite{duolingobuzz}. It also provides language certification
services\cite{duolingocert}.

Users are able to provide feedback on tasks, for example whether a translation
is reasonable or too hard at this proficiency level.

Duolingo tracks metrics, like what lessons people struggle the most with, 
adopting a wholesome data-driven approach. This combined with a flexible 
userbase allows them to iterate quickly on the assets provided in the 
lessons\cite{duolingodatadriven}.

But in terms of actually contributing with learning material, DuoLingo has 
 nothing to offer.

\subsubsection{Ease of use}

Duolingo appears to be easy to use. Quick feedback and positive reinforcement 
makes it easy to start using the system. There aren't many options available, 
except for audio settings (and other preferences), which makes the user 
interface easy to use.

\subsubsection{Takeaway points}

\begin{itemize}
\item Quick and simple feedback with additive gains facilitate user engagement
  greatly.
\item Community can be used to enhance or ensure the quality of lessons.
\item A data-driven approach might foster good community relations and provide more
  relevant, better assets as well.
\end{itemize}

\begin{CJK}{UTF8}{min}
\subsection{Evaluation of RPGツクール}
\end{CJK}

For something different, let's consider 
\begin{CJK}{UTF8}{min}RPGツクール\end{CJK} (RPG Tsuk\={u}ru). It's a game 
authoring tool wherein users create worlds through a tile-based map editor, 
and breathe life into their worlds by scripting actors, events and objects. 
Users may import their graphics and scripts, or use some of the wealth of art 
assets, scripted events and characters that are distributed with the program. 
The series is primarily successful in Japan, with English translations being a 
rather new thing.

While the main reason for us considering 
\begin{CJK}{UTF8}{min}RPGツクール\end{CJK} is its engaging user interface, it 
has in fact been used successfully for teaching. Authoring games in 
\begin{CJK}{UTF8}{min}RPGツクール\end{CJK} has proven successful in teaching 
mathematics\cite{maltempi2004learning}, and games authored in 
\begin{CJK}{UTF8}{min}RPGツクール\end{CJK} have been used for encouraging 
programming students\cite{Ralph_1the}. It is no secret that authoring and 
playing games is a sound way of both encouraging learners and increasing their 
efficacy.

\begin{CJK}{UTF8}{min}RPGツクール\end{CJK} assumes little or no experience in 
programming or designing games, and can as such be said to be aimed at 
newbies. However, several very successful commercial computer games by 
experienced developers were authored using 
\begin{CJK}{UTF8}{min}RPGツクール\end{CJK}.

\subsubsection{Expressiveness}

\begin{CJK}{UTF8}{min}RPGツクール\end{CJK} is primarily aimed at making 
Japanese Role-Playing Games such as the old 
\begin{CJK}{UTF8}{min}ファイナルファンタジ\end{CJK} (Fainaru Fantaj\={i}) or 
\begin{CJK}{UTF8}{min}イース\end{CJK} (\={I}su) games. The user may design 
tile-based 2D overhead-view maps with different types of graphics for the 
tiles, representing grass, ocean, and so on. Sprite sheets are used for 
animating characters. The user can script events and design a program flow 
through this.

Designing a map can in many ways be compared to using a bitmap editor. What 
You See Is What You Get. Authoring art assets outside of the ones that are 
distributed with the program is usually done with dedicated tools, such as 
using a video editor for making cut scenes, or an audio editor for making 
sound clips, or a bitmap editor for making sprite sheets. More recent versions 
have tools such as the character generator in which sprite sheets are actually 
generated by having the user choose settings such as the hair style and skin 
colour of the character they wish to generate.

The scripts for events and game objects in general are very powerful. The 
programming language Ruby is used in recent versions. Ruby is a fully-fledged 
Turing-complete programming language. The tool provides several high-level 
constructs to facilitate the scripting. Additionally there are tools for 
authoring scenarios and events through graphical user interfaces, comparable 
to the character generator.

\subsubsection{Ease of use}

The latest edition of \begin{CJK}{UTF8}{min}RPGツクール\end{CJK} boasts being 
``simple enough for a child; powerful enough for a 
developer''\cite{rpgmakervxace}. While certainly (theoretically) powerful, 
professional reviews are less than kind when discussing the user 
interface\cite{johnrpg} --- just like they've historically been unfavourable 
to the series\cite{gamespotrpg}.

However, interestingly, user reviews seems very positive on most 
sites\cite{metacriticrpg, amazonrpg, steamrpg}. And even more interestingly, 
most user reviews that explicitly mention the user interface are very happy 
with it, calling it ``clean'' and ``intuitive'' amongst other things.

The map editor is simple and intuitive to anyone who's ever used a bitmap 
editor of any kind. But the plethora of nested menus that needs to be endured 
to do anything else is unkind to the overall user experience. Add to that a 
poor overview, complex games quickly burdens the mind with the need to 
remember how your pieces all fit together, if they do so at all.

It may appear user reviews focus more on virtues of the map editor, whilst 
professional reviews can't get past the troubling interface for game objects 
in general.

In order to fully take advantage of the program, one arguably must learn the 
Ruby programming language to master scripts. Although we can hardly criticise 
\begin{CJK}{UTF8}{min}RPGツクール\end{CJK} for facilitating truly 
knowledgeable users, a better user-interface design might have bridged the gap 
to non-programmer users.

\subsubsection{Takeaway points}

\begin{itemize}
\item What You See Is What You Get is well-liked amongst users.
\item Horribly complicated nested menu systems are, uh, well, not well-liked 
    amongst professional reviewers\ldots
\item Authoring games can be a successful way of learning.
\item Playing games can successfully motivate learners.
\end{itemize}

\subsection{Evaluation of LateralGM and ENIGMA}

LateralGM\footnote{https://github.com/IsmAvatar/LateralGM} is a free tool to 
author and edit games. It aims to be compatible with the file formats of Game 
Maker and the newer GameMaker: Studio, and has a user interface very similar 
to that of Game Maker. To compile and run games, LateralGM normally interfaces 
with ENIGMA\footnote{http://enigma-dev.org/}, which is a compiler and runtime 
system aiming to be backwards compatible with Game Maker games.

\subsubsection{Expressiveness}
LateralGM and ENIGMA form a multi-genre game authoring tool with features that 
make it easy to use for beginners, but also powerful for advanced users. 
Although the scripting language has some support for 3D 
graphics\cite{enigma3d}, the software mainly focuses on 2D graphics and 
gameplay.

LateralGM presents the game as a set of different resources, including sounds, 
background images, sprites, scripts, fonts, objects, and rooms. Users may edit 
sounds and images within a game using external programs.

Objects may be programmed to act on various events. Events include the 
``step'' event (fired once in every update cycle, e.g.\ 30 times per second), 
user input, collision detection, timers, action-/script-triggered events, and 
many more\cite{lgmevents}. Each event can be programmed using either 
drag-and-drop actions or scripts. The set of drag-and-drop actions is 
extensive and includes basic flow control, and may also be extended with 
action libraries.

Each object can be set to use a possibly animated sprite, which causes each 
instance of the object to be drawn with that sprite. This sprite may also be 
used for collision detection. The ``draw'' event is also available for 
producing more advanced graphics. Objects may also be set as invisible, 
appearing in the room editor and possibly providing collision detection with 
the associated sprite, but not visible during gameplay.

The room editor allows users to graphically place instances of objects, either 
freely or in a configurable grid, inside a room as they will appear in the 
game. The user may define multiple layers of background and foreground images, 
either as whole images, possibly repeated or stretched, or in the form of 
placeable tiles, using the image as a tileset.

LateralGM loads and saves games in the proprietary file formats of various 
versions of Game Maker. These file formats are not officially documented, but 
have been reverse-engineered by LateralGM developers\cite{lgmformats}. While 
LateralGM and ENIGMA aim to be compatible with both current and earlier 
versions of Game Maker and GameMaker: Studio, several of these versions of 
Game Maker have incompatibilities between themselves. Not all features are 
implemented in LateralGM and ENIGMA, and development is struggling to catch up 
with later versions while still being compatible with earlier versions.

\subsubsection{Ease of use}
LateralGM lets users make simple games without writing a single line of code, 
using only drag-and-drop actions and graphically placing object instances in 
each room. Parametres to the actions may be given as scripting language 
expressions, providing a smooth learning curve from simple actions to the full 
scripting language.

\subsubsection{Takeaway points}

\begin{itemize}
\item Drag-and-drop can be used together with a written programming language 
    for a very smooth learning curve for non-programmers.
\item Trying to keep up compatibility with proprietary tools and undocumented 
    file formats is difficult and may take more development resources than 
    available.
\end{itemize}

\subsection{Evaluation of Minecraft}

Minecraft\footnote{\url{https://minecraft.net/}} is a modern open-world 
(sandbox) game. It allows the user to freely place, as well as craft 
(potentially new) blocks in the game. Thus, it is possible to build whatever 
imagination and time allows for in the game. The game is very popular, 
especially with the younger demographics, and has been noted for its ability 
to inspire creativity in its users\cite{minecraftign}.

Being a virtual world, and due to its popularity and game mechanics, it has been
suggested as a tool in primary and secondary education. Partly due to its 
potential for stimulating a sense of presence in the 
students\cite{coudrayminecraft}, which games in general are notoriously 
successful at.

In the area of situated learning, Minecraft tends to be classified as a
multi-user virtual environment, which is of particular interest to teachers
and students alike. But it is a category of situated learning which tends to 
suffer from some common issues, ranging from standardisation (can others 
extend it?) to time (how quickly is the user able to use the 
game?)\cite{dawley2014situated}. Due to its unique proposition of low cost, 
huge popularity, and innate creativity, Minecraft seems to offer a novel 
platform for learning to take place\cite{coudrayminecraft}.

\subsubsection{Expressiveness}

Minecraft is very expressive in allowing the user to roam freely, harvest 
resources, build structures, and craft new tools. As such, there are many 
modifications available to extend the game with new functionality, like energy 
systems, or new entities. There is currently no official application 
programming interface for developers of such extensions, and most of the 
extending is done using disassembly and patching the game locally.

Minecraft offers a primitive realisation of boolean digital circuits named
``redstone''. Redstone is a resource found in the game, and might in its 
harvested
state be placed on the ground to carry signals. Redstone carries either a
logical low (false) or high (true) signal, which can be (de)activated through 
a number of means, for example levers. Redstone can also be used in recipes to
craft new blocks like signal repeaters or switchable lights.

More notably, the redstone system in Minecraft is able to emulate the primitives
needed to construct any analogue circuit. Thus, people are able to construct
everything from simple systems that open doors, to full-on 8-bit
processors\cite{minecraftutube}. Thus, the redstone system offers an interesting
value-proposition: Mojang has succesfully fooled hundreds of thousands of 
children and adults alike into understanding primitive digital
circuits. By correspondence, they understand some set of boolean algebra, making
them capable programmers as well.

\subsubsection{Ease of use}

Minecraft is readily accessible, but has a steep learning 
curve\cite{minecraftign}. Systems like TooManyItems\cite{toomanyitems} makes 
the game slightly more accessible by making it easy to peruse recipes, aiding 
discovery. However, Minecraft itself provides little of a set path for playing 
the game, save a final boss fight, which is optional, and pretty much 
orthogonal to the ordinary game mechanics.

\subsubsection{Takeaway points}

\begin{itemize}
\item There are a number of issues with using virtual worlds in situated
  learning. Minecraft provides a reasonable solution to many of them.
\item Seemingly complicated topics (like circuits) can in fact be taught to
  children, given the right incentives 
\end{itemize}

\subsection{H5P}

H5P is a standard allowing for reusable, portable e-learning modules. It does 
so by providing plug-ins for Drupal\footnote{\url{https://www.drupal.org/}}, 
WordPress\footnote{\url{https://wordpress.org/}}, and 
Joomla\footnote{\url{https://www.joomla.org/}}, along with a file format 
specification for distribution, as well as standard \emph{content types} with 
built-in editors.

The value proposition of H5P is that it integrates nicely with existing
infrastructure like databases and view layers, enabling developers to focus on
developing their modules.\cite{h5pwhy} I.e., it reduces the ceremony needed
to get going. Moreover, its portability provides a much-wanted ease of
deployment for site owners, as well as a bigger market for developers.

Before deciding whether to use H5P we needed to evaluate five core issues.

\subsubsection{Are the abstractions offered by H5P powerful enough?}

At its very core, H5P offers a contract for libraries, which might function as
full-blown applications, stand-alone libraries, or application programming 
interfaces for any combination of the above. 

This is done in part by making the libraries document their semantics
(`semantics.json'), i.e.\ the properties (data structure) that are used to set
up the application. Each property is described by its name, as well as its type.
Different things have different semantic requirements: E.g., the
`list' (ordered group) semantic type requires the name of a single entity in it
(`entity'). This information is used by the generic editor to allow users to
easily customise the library.

H5P also provides a container format that manages ceremony, like loading
the appropriate CSS and JavaScript files, as well as providing useful metadata 
like the version and name of the library. This format includes a JavaScript 
class which libraries extend to hook into H5P. Applications are required to 
implement an `attach' method, which accepts the identity of the container in 
which to attach the module.

Of particular interest for us is the central event dispatcher 
(`H5P.EventDispatcher'), which is used to dispatch actions within 
H5P\cite{h5pdispatch}. Thus, like Elm's ports\cite{elmports} or Facebook 
React's Flux architecture/dispatcher\cite{fluxdispatch}, a synchronous 
interface for internal communication is provided, which might prove useful for 
multi-component systems.

Beyond this, H5P leaves the library developers plenty of room to expand. The 
H5P standard is pretty much ``open world'' and doesn't necessarily limit the 
structure of functionality of the JavaScript libraries it wraps, enabling 
developers to make arbitrarily complex software.

However, it should be noted that H5P still depends on professional
developers, and doesn't provide primitives that necessary make any sense for
ordinary users. Therefore, the standard per se doesn't preclude a more
user-oriented canvas system.

In conclusion, H5P doesn't itself provide the needed abstractions necessary, but
doesn't hinder them from being developed. We can use H5P for now.

\subsubsection{Are the abstractions high-level enough for our purposes?}

Given our emphasis on an intuitive user-interface for liberating the BOP, H5P 
quite frankly doesn't cut it for the time being. But even if H5P is not nice 
enough to use (at the moment) for non-technical users turned authors, it 
\emph{does} play nice with other software --- or, at the very least, it stays 
out of the way. This is \emph{very} desirable to us.

Once we have H5P modules ``up and running'', so to speak, there's nothing 
stopping us from moving ahead with other kinds of e-learning modules. There is 
furthermore nothing precluding H5P being improved in the future; we could even 
conceivably do that ourselves.

The honest truth is that H5P is not good enough for our intended target 
demographic. But it \emph{is} free software that integrates nicely and may be 
improved in the future. We'll take it.

\subsubsection{Can we generate valid H5P modules?}

Yes. Another short answer, but a more positive one at that.

As already noted, H5P doesn't limit any system wrapped in its container
format. Therefore, any functioning JavaScript system could in principle be
embedded within a H5P module, provided it only operates on a single node (due 
to the complexity and performance implications of coordinating DOM operations 
on overlapping elements).

\subsubsection{Can we import H5P modules?}

Again, yes! We can! We would need to make something that would be able to 
parse the semantic definitions of H5P modules, and we'd have to use this in 
some meaningful way. Effectively we should enhance the authoring tool as well 
to make it more appropriate for our target demographic.

\subsubsection{Can we offer a new way of composing H5P?}

H5P has no composition guarantees, and resorts to Design by Contract (DbC) for
composite systems.\cite{h5pdbc} This does nothing to ensure a uniform
communication standard, which ensures flexibility, but does not provide a
uniform interface. The specification itself admits that this is due to the lack
of interfaces in JavaScript. Thus, the contracts are more by politeness and not
by obligation, potentially creating weak implementations.

Therefore, there is obviously definite value to be provided by a system
providing more stringent, verifiable contracts between components, for example
using a more powerful type system than the one found in JavaScript. This might
in turn result in more reusable and robust applications. Also, because H5P 
effectively provides no composition contracts, this system might function as a 
sandbox and provide useful feedback should future versions of H5P seek to 
provide more formal requirements for composite modules.

One related problem is that H5P provides no standard for sub-module 
inheritance/reuse of content, as the system only has a notion of full H5P 
modules. We must figure out a way for several modules to refer to the same 
data.

\section{Conclusions and further work}
Our idea is a canvas for composing e-learning modules, initially targeting 
H5P. It should be intuitive to use, yet potent enough to make sufficiently 
refined modules. A simple user interface is vital to construct a ladder out 
from the BOP, whilst a capable set of features is important to attract users 
in general. Via gamification, users are encouraged to share their modules, and 
to improve the modules of others. We encourage a self-regulating community 
that promotes quality through a rating system for modules.

By authoring modules, the student becomes the teacher, which leads to more 
effective learning. Learning by teaching is a powerful concept that enhances 
the self-efficacy of students.

The canvas system is an important step towards our goal. However, much remains 
to be done. We need to achieve a cohesive authoring experience. Either by 
making our own modules, or by figuring out a way to make e-learning modules 
authoring agnostic and cohesive at the same time.

\end{multicols}
\bibliography{paper}

\begin{thebibliography}{10}

\bibitem{prahalad2009fortune}
Coimbatore~Krishna Prahalad.
\newblock {\em The {F}ortune at the {B}ottom of the {P}yramid, revised and
  updated 5th anniversary edition: Eradicating poverty through profits}.
\newblock FT Press, 2009.

\bibitem{brand2013crafting}
Jeffrey Brand and Shelley Kinash.
\newblock Crafting minds in {M}inecraft.
\newblock 2013.

\bibitem{cortese2005learning}
Claudio~G Cortese.
\newblock Learning through teaching.
\newblock {\em Management Learning}, 36(1):87--115, 2005.

\bibitem{gossen2012search}
Tatiana Gossen, Marcus Nitsche, and Andreas N{\"u}rnberger.
\newblock Search user interface design for children: Challenges and solutions.
\newblock In {\em EuroHCIR}, pages 59--62, 2012.

\bibitem{gordon1908otter}
CE~Gordon.
\newblock The otter in massachusetts.
\newblock {\em Science}, pages 772--775, 1908.

\bibitem{belanger2011review}
Michael Belanger, Nicole Clough, Nesime Askin, Luke Tan, and Carin Wittnich.
\newblock A review of violent or fatal otter attacks.
\newblock {\em IUCN Otter Specialist Group Bulletin}, 28(1):11--16, 2011.

\bibitem{inal2006children}
Yavuz Inal, Hatice Sancar, and Kursat Cagiltay.
\newblock Children’s avatar preferences and their personalities.
\newblock In {\em Society for Information Technology \& Teacher Education
  International Conference}, volume 2006 1, pages 4259--4266, 2006.

\bibitem{sadler2012virtual}
Randall Sadler.
\newblock Virtual worlds: An overview and pedagogical examination.
\newblock In {\em Bellaterra journal of teaching and learning language and
  literature}, volume~5, pages 0001--22, 2012.

\bibitem{hearthstone}
Ben Kuchera.
\newblock Blizzard silenced {H}earthstone players, and it made the game
  amazing.
\newblock
  \url{http://www.polygon.com/2014/4/18/5625802/hearthstone-chat-Blizzard},
  2014.

\bibitem{uuforskrift}
Kommunal og~moderniseringsdepartementet.
\newblock Forskrift om universell utforming av informasjons- og
  kommunikasjonsteknologiske (ikt)-løsninger.
\newblock 9, 2013.

\bibitem{educational}
{Free Software Foundation, Inc.}
\newblock Why educational institutions should use and teach free software.
\newblock \url{https://www.gnu.org/education/edu-why.html}, 2015.
\newblock Visited 2015-09-24.

\bibitem{czaplicki2012elm}
Evan Czaplicki.
\newblock Elm: Concurrent frp for functional guis.
\newblock {\em Senior thesis, Harvard University}, 2012.

\bibitem{smitsextreme}
Jeff Smits.
\newblock Extreme pong: A browser-based game implemented using functional
  reactive programming.

\bibitem{flanagan2006javascript}
David Flanagan.
\newblock {\em JavaScript: the definitive guide}.
\newblock " O'Reilly Media, Inc.", 2006.

\bibitem{buist2014extending}
Simon Buist.
\newblock Extending an ide to support input device logging of programmers
  during the activity of user-interface programming: Analysing cognitive load.
\newblock {\em Bachelor of Science dissertation, The University of Bath}, 2014.

\bibitem{berntsen2014quest}
Alexander Berntsen.
\newblock The quest for programming {N}irvana: On programming game systems in
  {H}askell.
\newblock 2014.

\bibitem{kraeutmann2015functional}
David Kraeutmann and Philip Kindermann.
\newblock Functional reactive programming and its application in functional
  game programming.
\newblock 2015.

\bibitem{elmports}
Elm.
\newblock Interop.
\newblock \url{http://elm-lang.org/guide/interop}.
\newblock Visited 2015-09-16.

\bibitem{marlow2010haskell}
Simon Marlow et~al.
\newblock Haskell 2010 language report.
\newblock \url{http://www.haskell.org/onlinereport/haskell2010/}, 2010.

\bibitem{hughes1989functional}
John Hughes.
\newblock Why functional programming matters.
\newblock {\em The computer journal}, 32(2):98--107, 1989.

\bibitem{wadler1989make}
Philip Wadler and Stephen Blott.
\newblock How to make ad-hoc polymorphism less ad hoc.
\newblock In {\em Proceedings of the 16th ACM SIGPLAN-SIGACT symposium on
  Principles of programming languages}, pages 60--76. ACM, 1989.

\bibitem{jones2003wearing}
Simon~Peyton Jones.
\newblock Wearing the hair shirt: a retrospective on haskell.
\newblock {\em Invited talk at POPL}, 206, 2003.

\bibitem{jones2012future}
Simon~Peyton Jones.
\newblock The future is parallel, the future of parallel is declarative.
\newblock 2012.

\bibitem{claessen2011quickcheck}
Koen Claessen and John Hughes.
\newblock Quick{C}heck: a lightweight tool for random testing of {H}askell
  programs.
\newblock {\em Acm sigplan notices}, 46(4):53--64, 2011.

\bibitem{arts2006testing}
Thomas Arts, John Hughes, Joakim Johansson, and Ulf Wiger.
\newblock Testing telecoms software with quviq quickcheck.
\newblock In {\em Proceedings of the 2006 ACM SIGPLAN workshop on Erlang},
  pages 2--10. ACM, 2006.

\bibitem{memrise}
Memrise.
\newblock Memrise science.
\newblock \url{http://www.memrise.com/science/}.
\newblock Visited 2015-10-09.

\bibitem{potts2014benefit}
Rosalind Potts and David~R Shanks.
\newblock The benefit of generating errors during learning.
\newblock {\em Journal of Experimental Psychology: General}, 143(2):644, 2014.

\bibitem{kidsquora}
What's important when designing a touch user interface for kids apps?
\newblock
  \url{http://www.quora.com/Whats-important-when-designing-a-touch-user-interface-for-kids-apps}.

\bibitem{kidsluke}
Luke Wroblewski.
\newblock Touch-based app design for toddlers.
\newblock \url{http://www.lukew.com/ff/entry.asp?1179}, 2010.

\bibitem{ankimanual}
Damien Elmes.
\newblock Anki 2.0 user manual.
\newblock \url{http://ankisrs.net/docs/manual.html}.
\newblock Visited 2015-10-09.

\bibitem{pcworldanki}
Erez Zukerman.
\newblock Review: Anki helps you to learn and memorize virtually anything, for
  free.
\newblock
  \url{http://www.pcworld.com/article/2030095/review-anki-helps-you-to-learn-and-memorize-virtually-anything-for-free.html}.

\bibitem{burke2002s}
Jim Burke.
\newblock It's all in the cards.
\newblock {\em Voices from the Middle}, 10(1):54, 2002.

\bibitem{benassi2014applying}
Victor Benassi, Chris Hakala, Sarah~A Ambrose, Marsha Lovett, Courtney~M Clark,
  Robert~A Bjork, Chee~Ha Lee, Slava Kalyuga, John~AC Hattie, Gregory~CR Yates,
  et~al.
\newblock {\em Applying science of learning in education: Infusing
  psychological science into the curriculum}.
\newblock 2014.

\bibitem{duolingobuzz}
Luis.
\newblock Duolingo now translating buzzfeed and cnn.
\newblock \url{https://www.duolingo.com/comment/954969}, 2014.
\newblock Visited 2015-10-09.

\bibitem{duolingocert}
Burr Settles.
\newblock Duolingo launches its certification program to take on toefl.
\newblock
  \url{http://techcrunch.com/2014/07/23/duolingo-launches-its-language-certification-program/},
  2013.

\bibitem{duolingodatadriven}
Burr Settles.
\newblock Duolingo’s data-driven approach to education.
\newblock
  \url{http://duolingo.tumblr.com/post/41960192602/duolingos-data-driven-approach-to-education},
  2013.

\bibitem{maltempi2004learning}
Marcus~Vinicius Maltempi and Maur{\'\i}cio Rosa.
\newblock Learning vortex, games and technologies: a new approach to the
  teaching of mathematics.
\newblock In {\em INTERNATIONAL CONGRESS ON MATHEMATICAL EDUCATION}, volume~10.
  Citeseer, 2004.

\bibitem{Ralph_1the}
Tiffany Ralph and Tiffany Barnes.
\newblock 1 the catacombs: A study on the usability of games to teach
  introductory programming.

\bibitem{rpgmakervxace}
{RpgMaker.org}.
\newblock {RPG} {M}aker {VX} ace.
\newblock \url{http://www.rpgmaker.org/downloads/rpg-maker-vx-ace-downloads/},
  2015.
\newblock Visited 2015-10-09.

\bibitem{johnrpg}
{John Wizard II}.
\newblock John {W}izard {R}eview {RPG} {M}aker {VX} {A}ce.
\newblock \url{http://www.johnwizard.com/rpg-maker-vx-ace-review.php}, 2015.

\bibitem{gamespotrpg}
{Miguel Lopez}.
\newblock Rpg maker review.
\newblock \url{http://www.gamespot.com/reviews/rpg-maker-review/1900-2633694/},
  2000.

\bibitem{metacriticrpg}
\url{http://www.amazon.com/RPG-Maker-VX-Ace-Pc/product-reviews/B00AA25QUM}.
\newblock Visited 2015-08-31.

\bibitem{amazonrpg}
\url{http://www.amazon.com/RPG-Maker-VX-Ace-Pc/product-reviews/B00AA25QUM}.
\newblock Visited 2015-08-31.

\bibitem{steamrpg}
\url{https://steamcommunity.com/app/220700/reviews/}.
\newblock Visited 2015-08-31.

\bibitem{enigma3d}
3d graphics functions.
\newblock \url{http://enigma-dev.org/docs/Wiki/3D_Graphics_Functions}.
\newblock Visited 2015-09-01.

\bibitem{lgmevents}
Quadduc Clam, IsmAvatar and Robert~B. Colton.
\newblock Event.java.
\newblock
  \url{https://github.com/IsmAvatar/LateralGM/blob/dc99fbf54e586d435ad488879b399bd2df362210/org/lateralgm/resources/sub/Event.java}.
\newblock Visited 2015-09-01.

\bibitem{lgmformats}
\url{http://sourceforge.net/projects/lateralgm/files/Formats/GM/}.
\newblock Visited 2015-09-01.

\bibitem{minecraftign}
Anthony Gallegos.
\newblock Minecraft review.
\newblock \url{http://uk.ign.com/articles/2011/11/24/minecraft-review}, 2011.
\newblock Visited 2015-08-31.

\bibitem{coudrayminecraft}
Catherine Coudray.
\newblock Minecraft in higher education, beyond learning activities.

\bibitem{dawley2014situated}
Lisa Dawley and Chris Dede.
\newblock Situated learning in virtual worlds and immersive simulations.
\newblock In {\em Handbook of research on educational communications and
  technology}, pages 723--734. Springer, 2014.

\bibitem{minecraftutube}
William~Wyatt Earnshaw.
\newblock Minecraft reprogrammable 8-bit redstone computer.
\newblock \url{https://www.youtube.com/watch?v=nlt0xMVoJcM}, 2013.

\bibitem{toomanyitems}
\url{http://www.minecraftforum.net/forums/mapping-and-modding/minecraft-mods/1272385-toomanyitems-the-inventory-editor-and-more-1-8}.
\newblock Visited 2015-08-31.

\bibitem{h5pwhy}
H5P.
\newblock Why develop for h5p?
\newblock \url{https://h5p.org/node/3150}, 2015.
\newblock Visited 2015-09-10.

\bibitem{h5pdispatch}
H5P.
\newblock H5p api reference.
\newblock \url{https://h5p.org/documentation/api/H5P.EventDispatcher.html},
  2015.
\newblock Visited 2015-09-10.

\bibitem{fluxdispatch}
Bill Fisher.
\newblock Flux: {A}ctions and the {D}ispatcher.
\newblock
  \url{https://facebook.github.io/react/blog/2014/07/30/flux-actions-and-the-dispatcher.html},
  2014.

\bibitem{h5pdbc}
H5P.
\newblock Contracts.
\newblock
  \url{https://h5p.org/documentation/for-developers/contracts-and-addons},
  2015.
\newblock Visited 2015-09-10.

\end{thebibliography}
\bibliographystyle{unsrt}
\newpage
\appendix
\section{Design document}
\label{design}
This is the design document authored for our system. It is authored by a user 
experience expert, Lene Thuseth, as we felt the interface emphasis warranted 
professional involvement.

While it certainly does not cover all of our ideas, it does showcase the core 
subset of what we envision in a minimum viable product. More importantly, it 
brings our user-interface research observations to life, and shows the 
fundamental interaction pattern and establishes the ever intangible 
``look'n'feel'' of our system. The shapes, colours, fonts, and overall 
presentation is given careful thought. This is all described and explained in 
the document.



\section*{Supplementary material}

# DESIGN DOCUMENTATION

**Enabling Learning By Teaching:**
Intuitive Composing of E-Learning Modules

21st of October 2015

Author: Lene Thuseth

# CONTENTS

# THE IDEA

The idea is to make a highly intuitive and engaging system to compose e-learning modules. The system will support authoring of both linear and non-linear e-Learning[1], and is designed with an emphasis on younger users. In the MVP (Minimum Viable Product) the system will allow users to compose H5P-modules (h5p.org) with an unmodified version of the H5P composing tools.

The design and concept is based on research done and published by plaimi (secure.plaimi.net). More information on market research, target audience and the concept itself can be found in their paper[2].

> *"In an effort to foster learning by teaching, we propose the development of a canvas system that makes composing e-learning modules intuitive. We try to empower and liberate non-technical module users by lowering the bar for turning them into module authors, a bar previously set far too high. In turn, this stimulates learning through teaching. By making a damn fine piece of software, we furthermore make module authoring more pleasant for experienced authors as well. We propose a system that initially enables users to easily compose H5P modules. These modules are successively easy to share and modify. Through gamification we encourage authors to share their work, and to improve the works of others."*
>
> Abstract from the Enabling Learning by Teaching: Intuitive Composing of E-Learning Modules paper

The focus of the concept is to foster "learning by teaching", by encouraging students to make and share e-learning modules, and re-use and re-mix modules made by others. At the same time we make it much easier for professional e-learning authors to make engaging and adaptive e-learning.

The design itself will have a heavy focus on all aspects of usability, including universal design (accessibility), motivation by gamification, and how intuitive and easy it is to use and navigate the system.

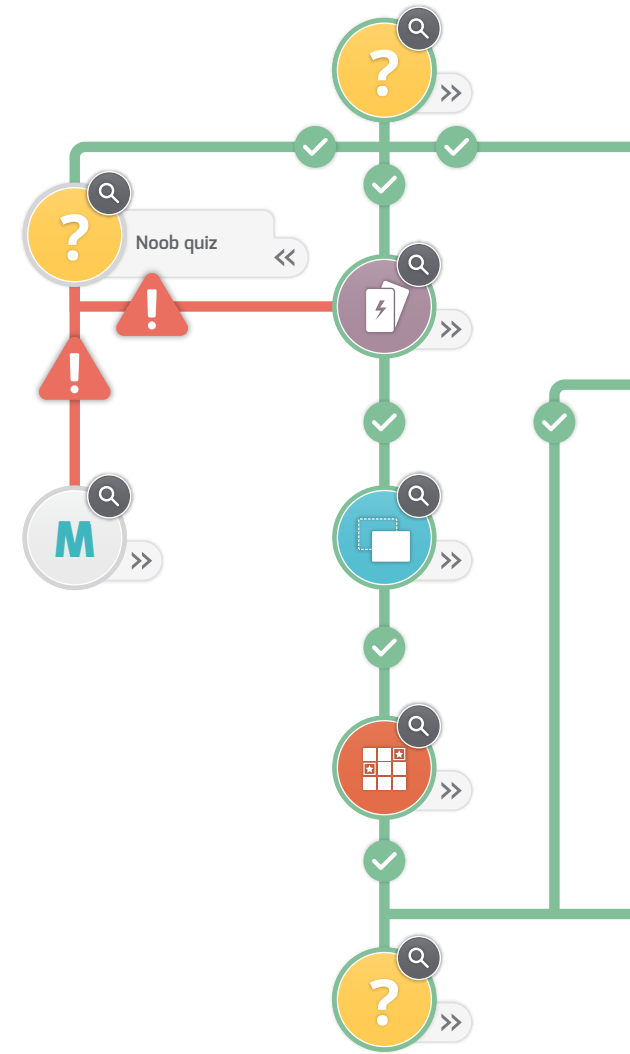

# THE DESIGN

## The look

Because children are the primary target audience, the design makes use of bright colours, strong contrasts, big buttons, playful avatars, and a friendly language. At the same time kids aren't the *only* target audience, so the design also needs to appeal to a more mature audience. By using bright but slightly muted colours we achieve a colour palette that looks fun and child-friendly but not too childish. This also gives it a modern feel.

Round shapes such as circles and rounded corners feel friendly and playful, enticing the user to play with the system, and that is exactly the message we want the design to send; "come play with me, come have fun!"

Big buttons and strong contrasts also help the user navigate the system, by making it obvious how to interact and where to click. By having a tidy interface with only the necessary functionality available, we avoid confusion and frustrating the user. The less time the user spends trying to figure out how to use the system, the more time they can spend on the actual task.

Buttons that can be clicked use visual effects such as gradients and shadows to make them appear elevated from the surface. People intuitively know that they can click these buttons.

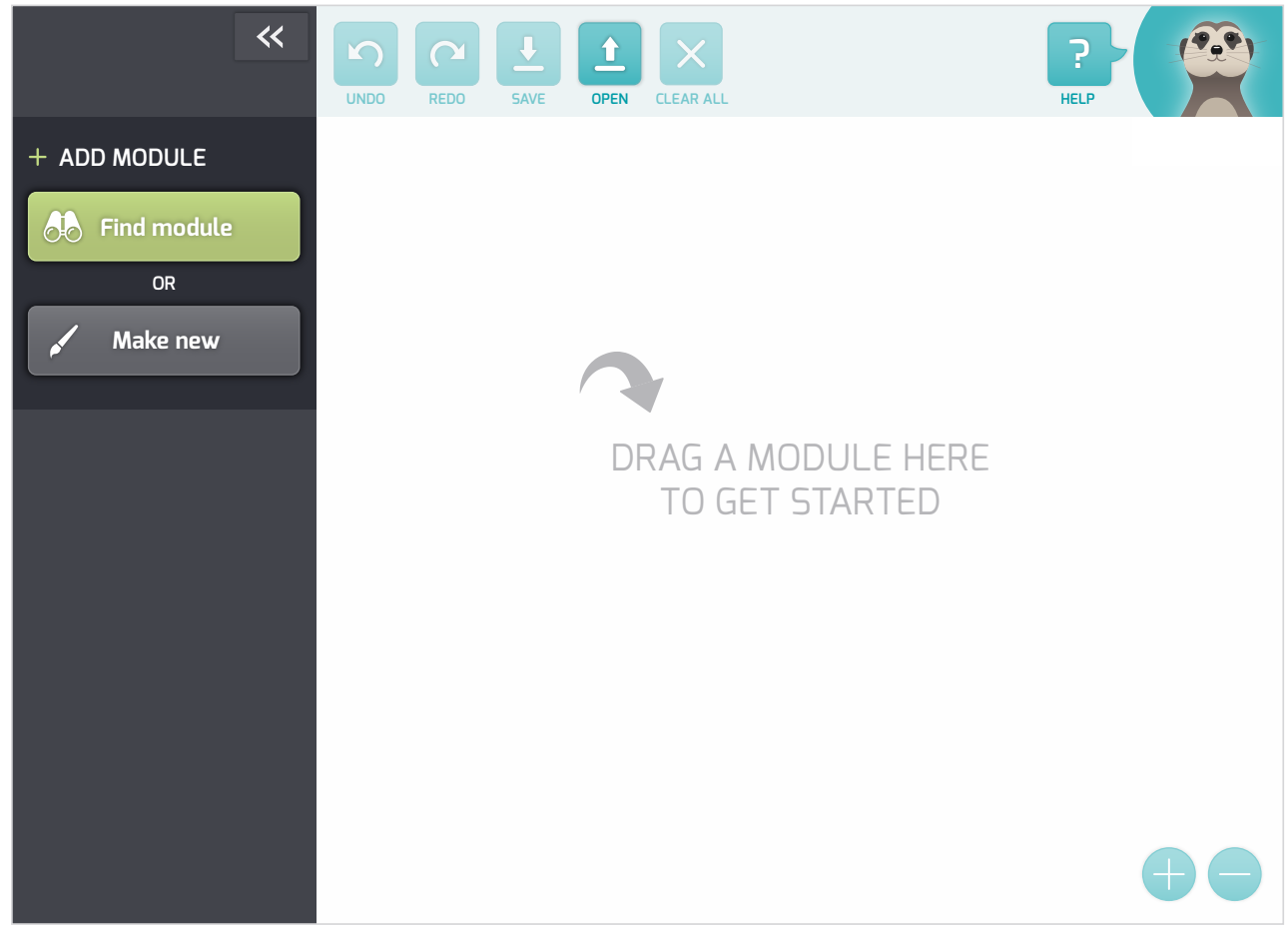

## The language

By using a casual and friendly language we try to introduce a sense of humanity and human touch to the system. It feels like your friend rather than a cold and robotic system. We make use of an avatar that will pop up with hints to help the user along, or to let them know when they've done something wrong. We don't want to give kids a feeling they've done something wrong, so even when there's an error message it is phrased in a positive and friendly manner.

Instructions on buttons and actions are clear and concise. We use the terminology *find* rather than *search*, and well-established terms such as *undo* and *save*.

The language aims to be fun and easy to understand for kids, but clear and comprehensible for adults.

We use metaphors to make something a little bit complex, such as flows and forks, more understandable. Since the avatar is an otter, we use the metaphor of rivers and dams when we talk about the flow through the modules, and possible broken logic that impedes the flow. This can help make a complex concept seem much more understandable for kids as well as adults, and help them author complex e-learning modules/courses.

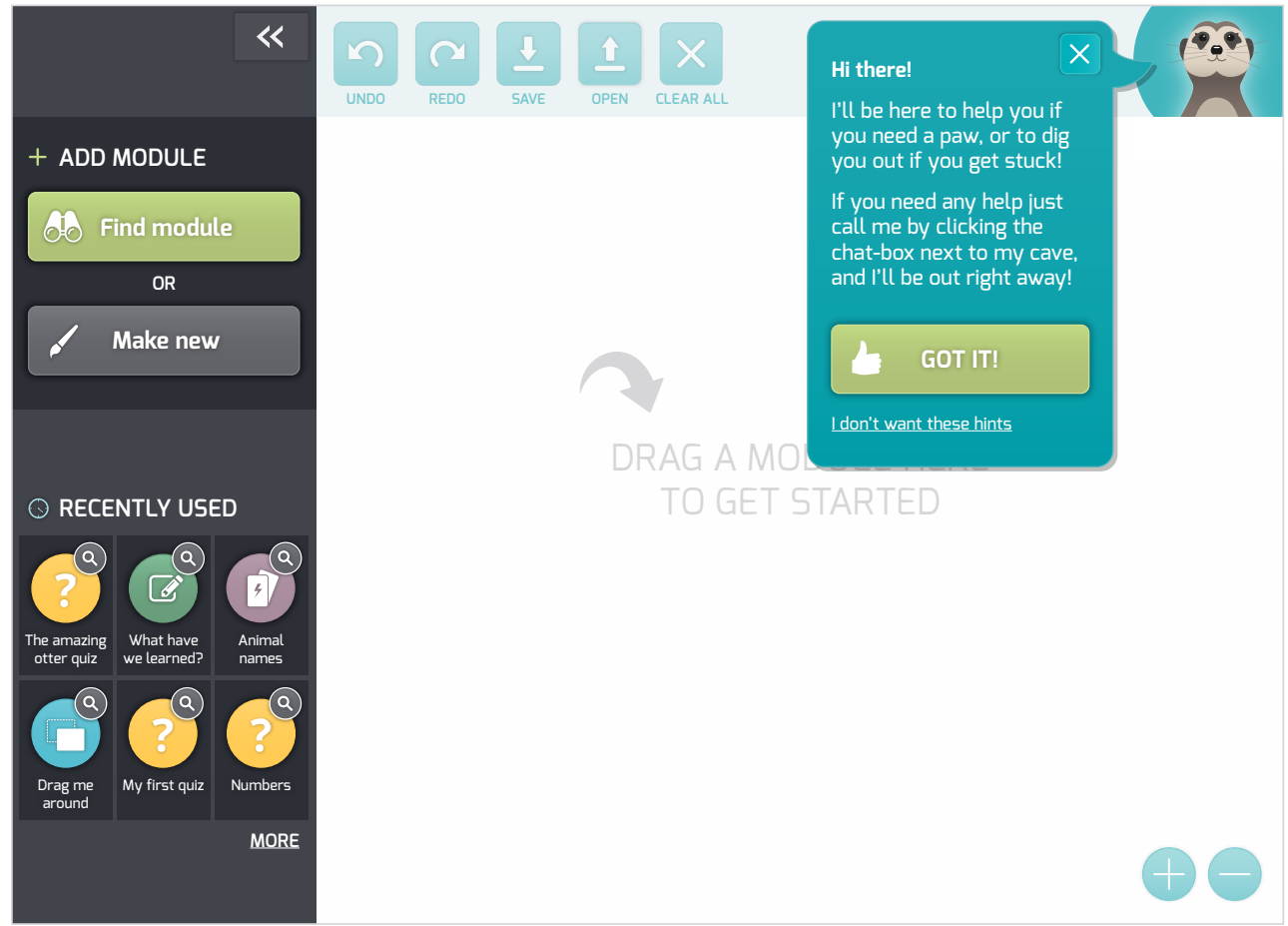

## Functionality and responsiveness

The most important thing of a system of any kind is its functionality. It doesn't matter how beautiful it looks if it doesn't feel responsive and logical. It needs to give prompt feedback when actions are performed, so the system feels alive and responsive.

We have no splash screens or loading bars. Actions lead to an instant response, e.g. if a user searches for something, the search results populate as they type giving instant feedback to their actions. Kids, especially, hate waiting, and will not have the patience for fancy splash screens. Therefore the user may jump straight into the task with no delay as soon as they enter the system.

Our focus on responsiveness is further emphasised by things like hover effects for mouse-users and on-tap effects for tablet users. When you hover a button, its colour will change, and it will animate (e.g wiggle) in order to convey a possible interaction. Furthermore, when the button is clicked, its appearance changes as a visual feedback. As an example the 'CONNECTIONS' actions are in a toggle bar where either one or fewer of the actions can be active. When one action is activated the mouse cursor changes to indicate the action that can be performed, e.g. the 'DRAW' function transforms the cursor into a pencil.

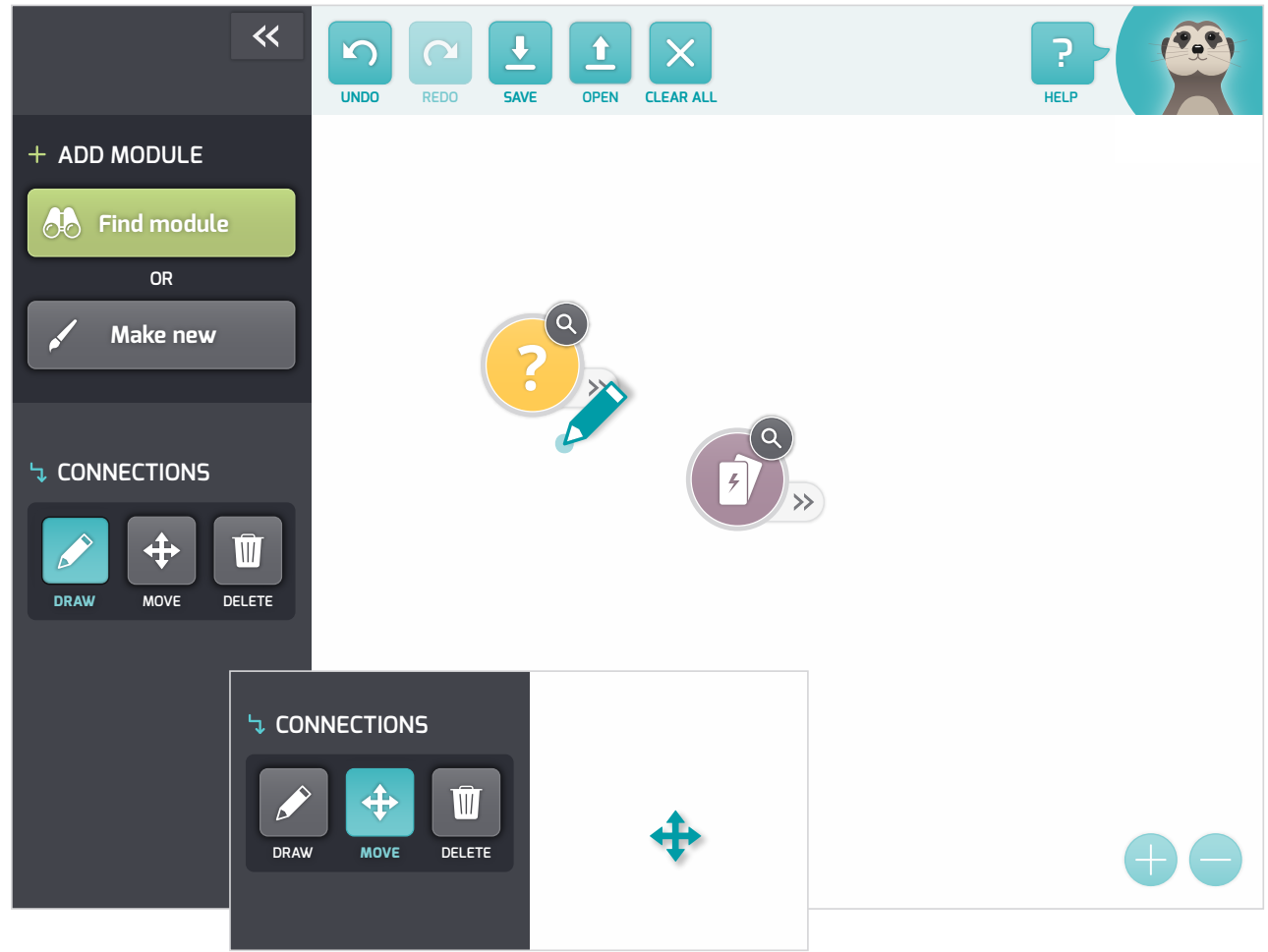

## Functionality and responsiveness (cont.)

The more complex ways the system is responsive and feel alive, is by understanding a user's behaviour. If they fail at a task or take too long to do something, the system will assume the user might need help with what to do. One example is if the user starts the system and doesn't do anything within a certain time limit, the avatar will pop up with a hint on how to get started. If a user starts performing actions right away, this hint will not appear. The same hints will not appear more than once.

Another example is if the user breaks the logic of a connection. The hint will pop up telling them that their latest changes has caused errors in their flow, and how to fix it.

All 'big' actions such as 'CLEAR ALL' will have an extra layer of confirmation before the action is completed in case of enabled by accident. All actions can be reversed with the undo function or re-done with the redo function.

We never use just icons for important functionality, but include text labels as well. The only exception is when the narrow side bar is active by choice of the user.

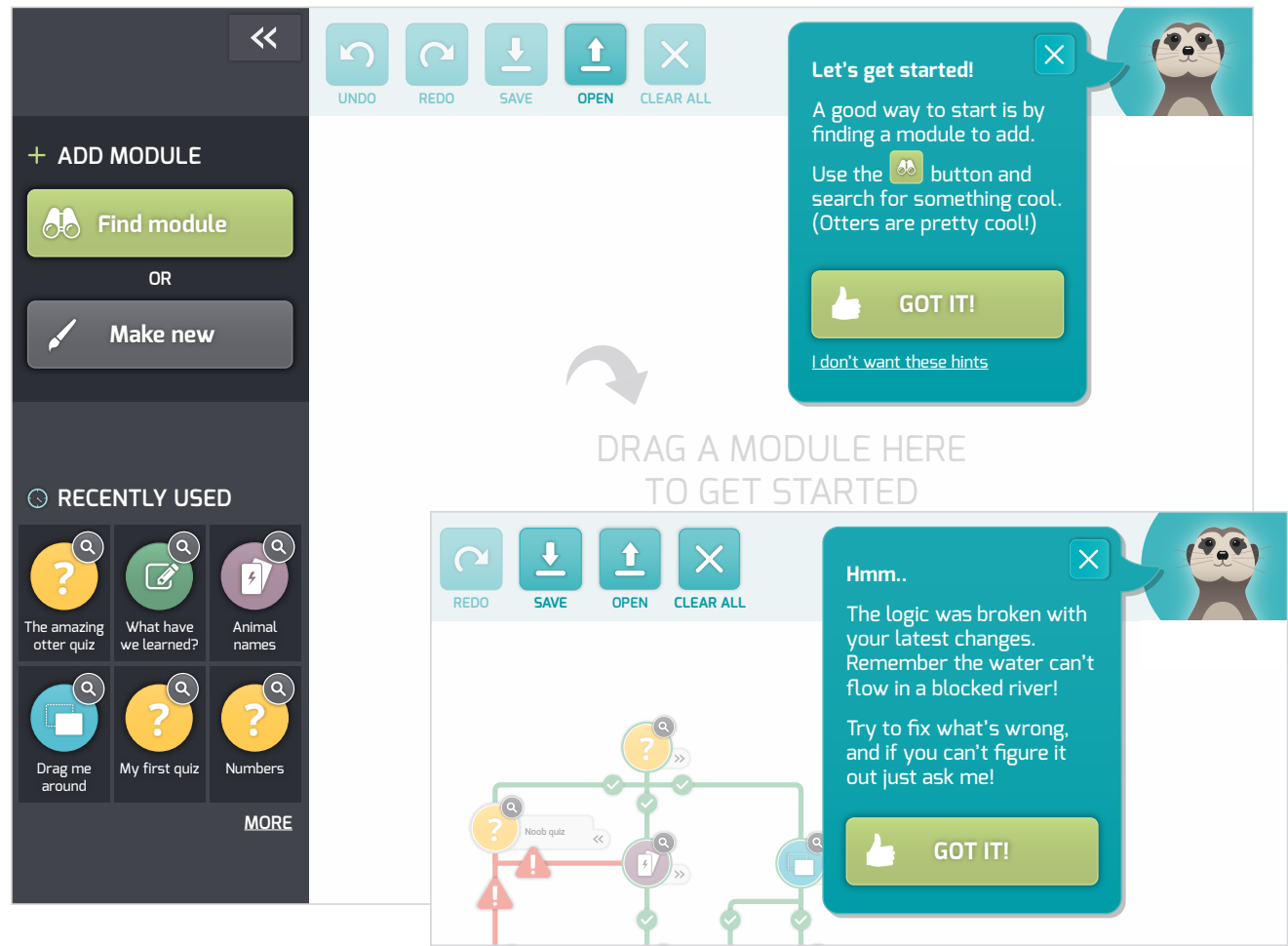

## Consistency

To improve a system's usability, the navigation needs to be consistent. If a button does something different in one place in a system than it does in other places, the user may get confused and frustrated. Consistency improves predictability, which in turn increases learnability[3] (how quickly a user learns how to use the system).

This system has very consistent navigation. An example of this is how the module icon can be interacted with. If the module icon or magnifying glass icon is selected, the info box will pop up. This action is available no matter where the icon appears. The magnifying glass icon indicates that the module isn't empty, unlike when the icon is on its own when adding a *new* module. As soon as the module is authored, the magnifying glass is added to show that the module is no longer empty. From this info box, the module can be 'EDITED' (re-mixed), 'ADDED' (re-used) if it is not on the canvas already or 'DELETED' (from the canvas). So the actions are contextual, but the navigation is consistent and can therefore be predicted and expected by the user. The actions appear in order of importance, 'ADD' is more important than 'EDIT' (as it allows for quicker use), and 'EDIT' is more important than 'DELETE'.

There are also colour consistencies in buttons across the system with similar actions.

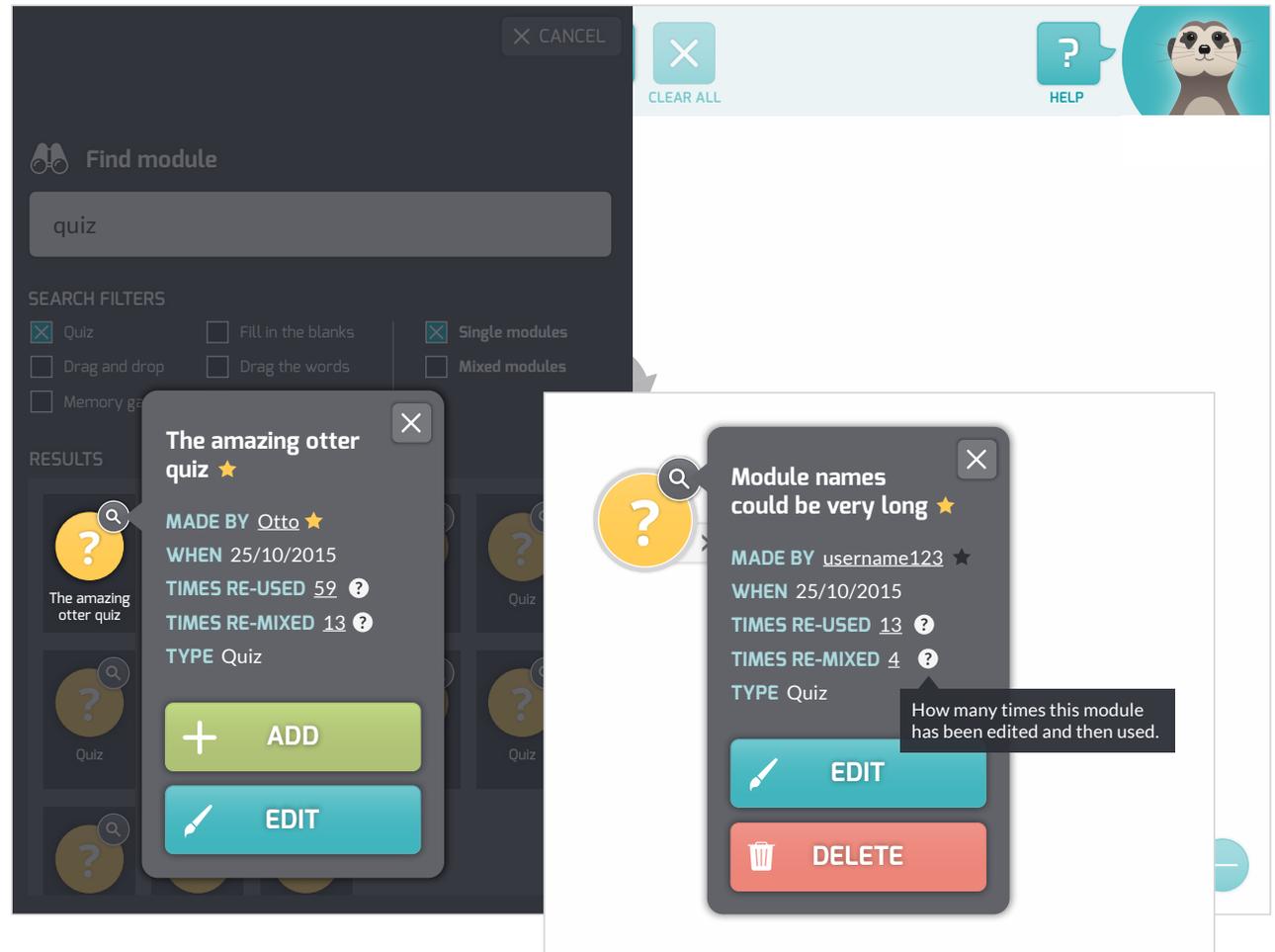

## Animations and transitions

Another way to enhance the feeling of consistency is by applying animations when transitioning to a new page by clicking a button. A context shift can cause unease if it's not visually expected. By using animations, such as morphing, we can ease the transition for the user.

As an example we have the two buttons in the toolbox default view; 'Find module' and 'Make new'. When clicked these will transition into respectively the search page and the make new page. We ease context-switching with animations. E.g. by morphing the search button into its new state, and leaving it active for the user to type into straight away, we create a much less confusing context-switch.

Similarly, the two buttons will move and morph into their current appearance on the "Make new" page rather than just appearing in their new states.

By selecting the cancel buttons on each of the new screens, it will morph back into its previous state.

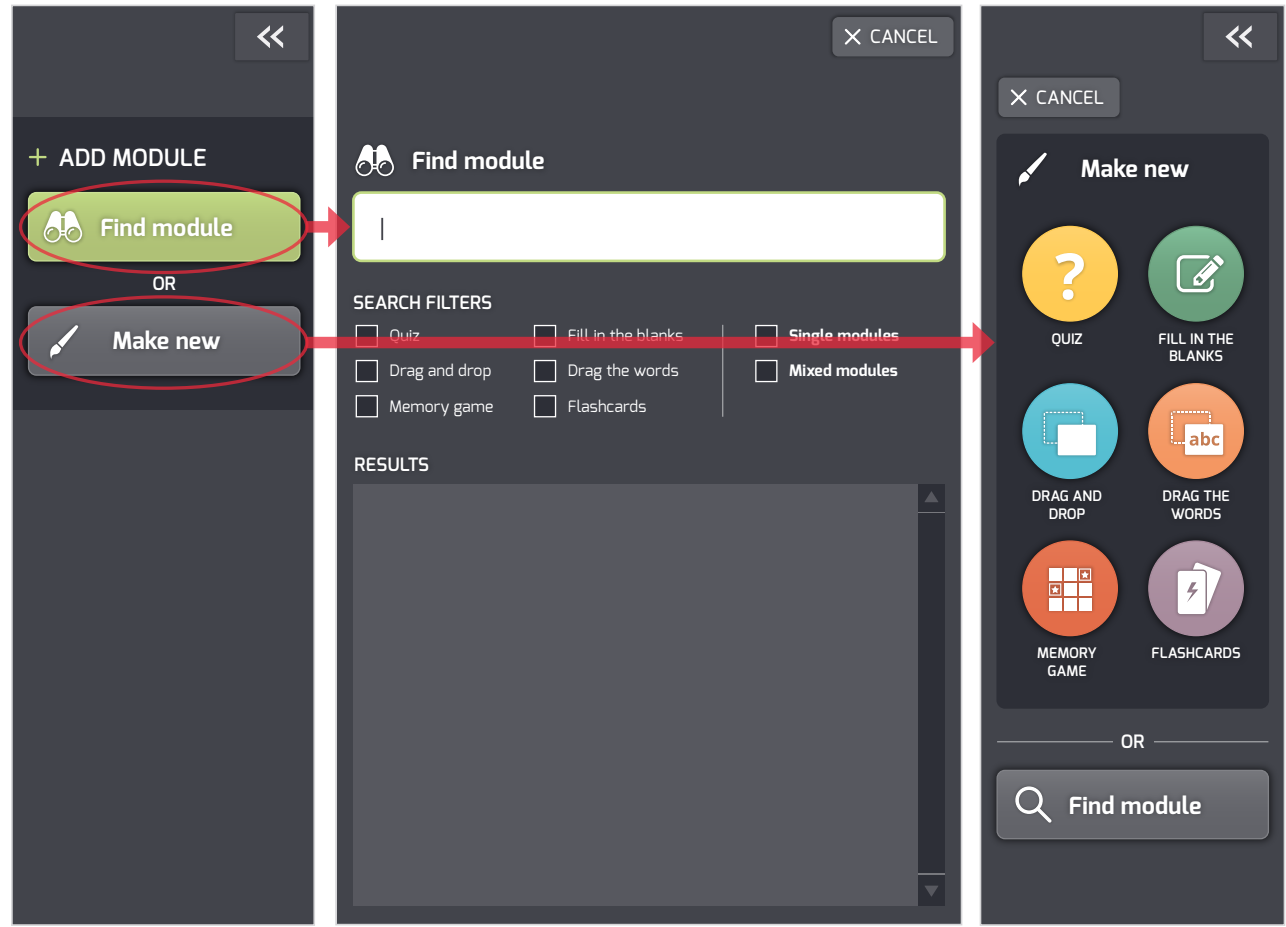

## Gamification

Gamification is when you apply game mechanics and metaphors to non-game scenarios in order to motivate and engage people to reach their goals.

Our system will be gamified, with the use of customisable avatars and achievements, and with the look and feel of a game, by allowing the user to drag items around freely to accomplish their goals.

We use gamification, especially to motivate learners to share and re-use other people's modules. We do this by awarding achievements for re-using modules, and for having modules that are frequently re-used. To prevent meta-gaming* we can add an 'influence' system, wherein a module gets a higher score if used by more unique users. The influence system influences module popularity.

Users may customise their avatar, which gives the feeling of having their own personal helper. This creates more attachment and loyalty to the system, and makes it feel more like a friend than a computer system. Gamification elements like these make the user feel more involved, and in turn more engaged in the task that they are doing.

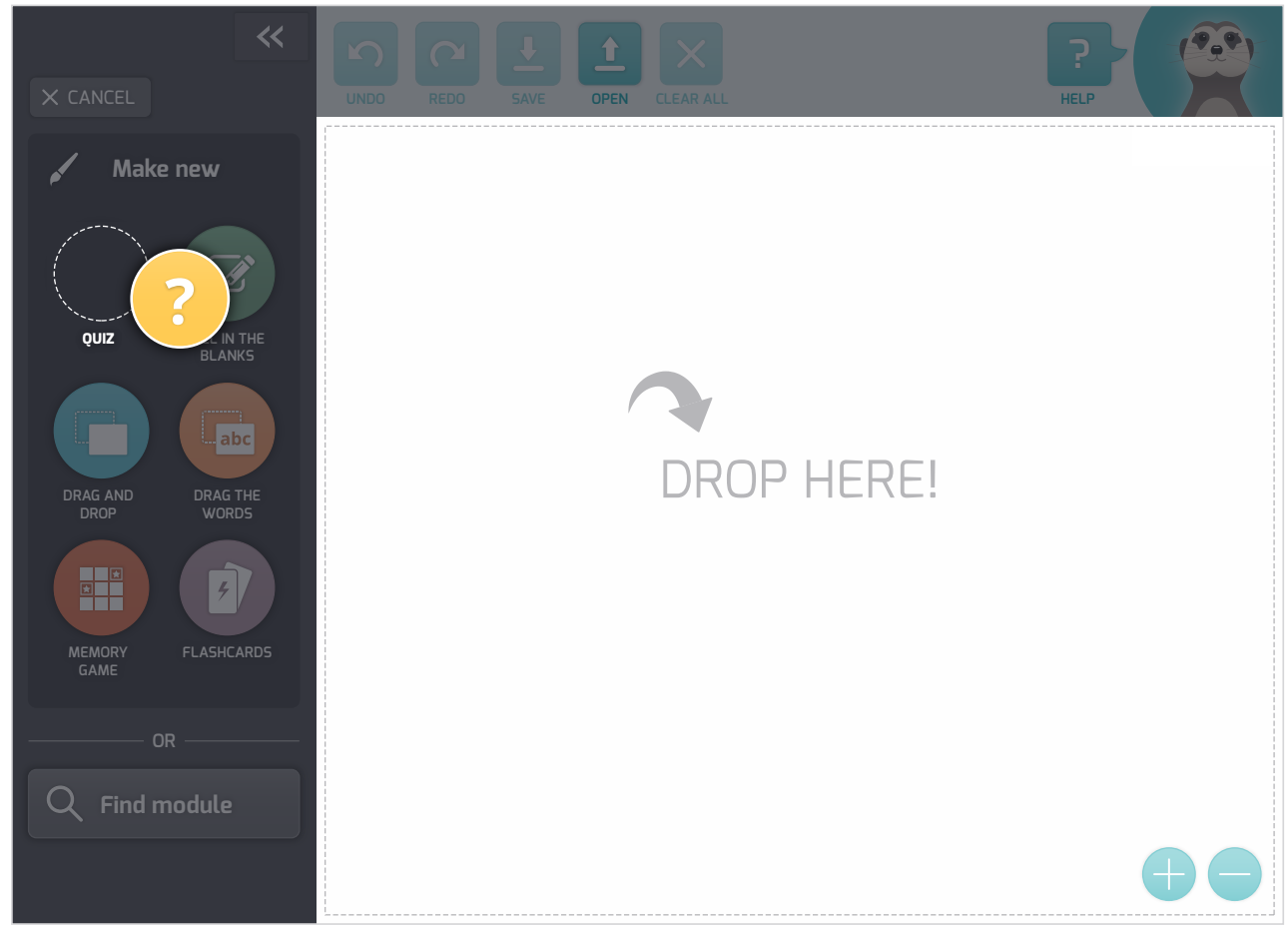

*Exploiting the game mechanics in order to achieve a high score.

## Universal design (accessibility)

Our system will be a website. One of the main motivating factors for this decision is that we want our system to be available to as many people as possible. This means we need to support the most popular operating systems and devices. This also means we need to take into account accessibility software such as screen-readers. Our system aims to follow the WCAG2.0 requirements as outlined by Difi. [(uu.difi.no)](uu.difi.no)

We do this in several ways. One way is making sure all colours have a high enough contrast ratio, and that we don't use colour by itself to indicate things. An example is the connections, which appear red or green, depending on whether the logic is correct or not, that also have an icon in a different shape to indicate the same thing.

The primary way of navigating the system relies on drag-and-drop, but we also have alternative methods of navigating. E.g. the icons may be dragged to the canvas, or they may be selected, at which point an 'ADD' button will appear. Selecting this button is functionally equivalent to the drag-and-drop action.

We use large buttons with larger hit-boxes because children are in the target demographic. E.g. the module-button: activating the module icon does the same thing as activating the magnifying glass. The magnifying glass serves primarily as a visual indication.

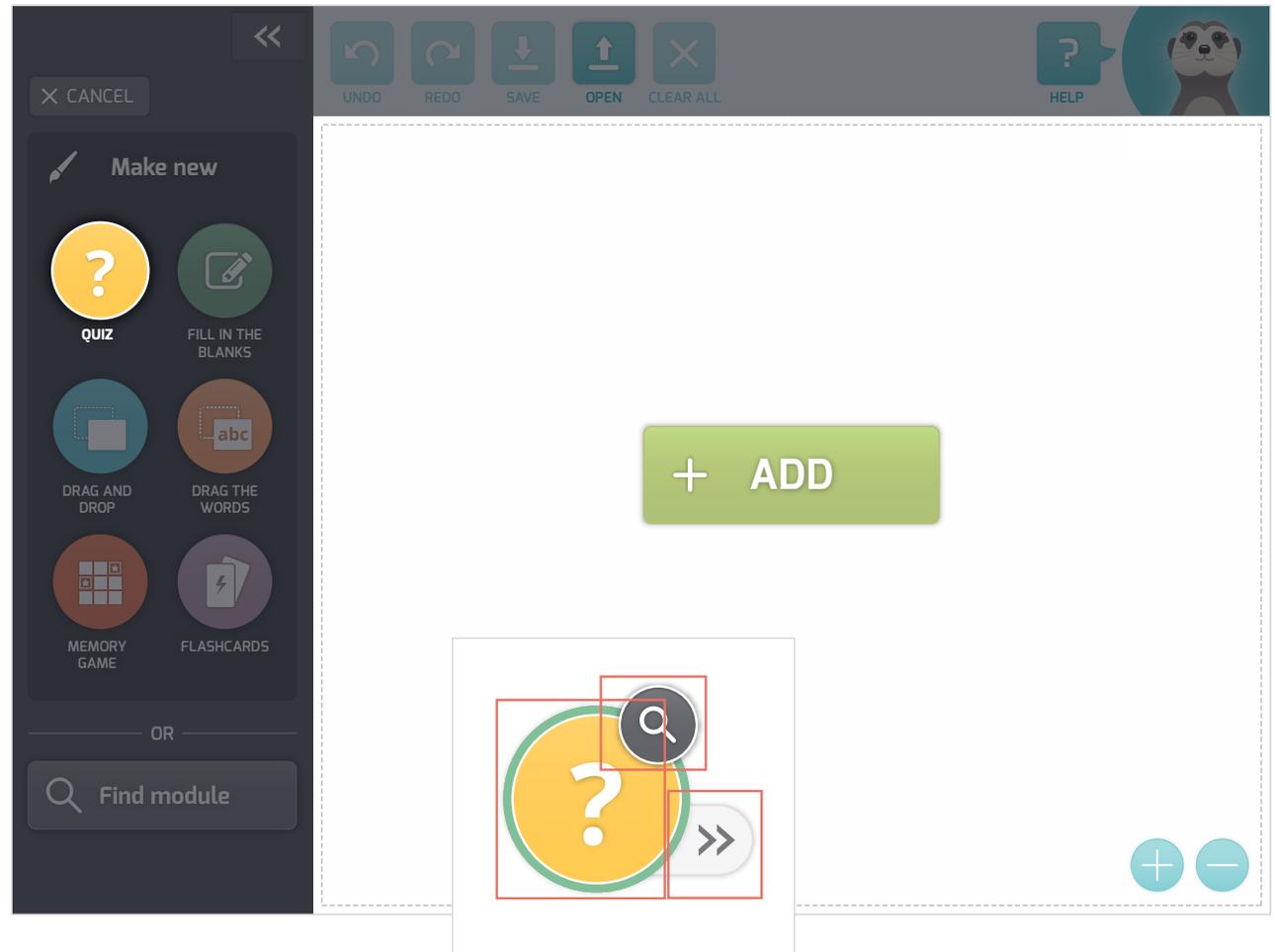

# THE CONCEPT

## The essence

What makes this product so unique, is that it makes it easy for anyone, regardless of age, background, or technical ability, to author sophisticated non-linear module structures. This is something that previously has only been available to professionals working in complex systems. Our system is so intuitive and easy to use that the user can focus almost entirely on the task at hand.

The biggest drawback at this stage is the complicated and not very user-friendly H5P authoring system. However, this can be seen as an opportunity to work on the user experience of the H5P-system. This will in turn benefit H5P itself and every user of it, including our system. Additionally, the authoring capabilities of future e-learning modules might be designed to support our system from the ground up.

Research done in the earlier stages of this project indicate that there are several operators in the market that have some of the same features that this project has, but none that do exactly what we aim to do. There are many module authoring systems, and many e-learning systems for kids, as well as several canvas systems where users can design their own games or experiences. But there are none focussing on visual and intuitive authoring of e-learning modules. Nor any focussing on the learning by teaching

effect. By using gamification to encourage students to share, use, and improve other people's work, we allow for a growth culture for social learning. We encourage problem-solving via gamification. The user is free to move items around on their "canvas" as they please, and visually build up their course, with actions and consequences. A student can easily author a non-linear module by using if-expressions. E.g. if the score of a module is less than 80% then go to this module, if not proceed to the next module.

The system helps the user formulate if-expressions. E.g. the user has four modules, A, B, C, and D. They connect B to A with the condition of a 50% completion score of A. They then connect C to A with a 75% completion score condition. Finally they connect D to A. D then automatically serves as a "catch all", by catching all scores less than 50%. The users are of course allowed to overrule this, but then they also have to fix the broken logic.

A good way to encourage kids to learn is to involve them in their learning, and a great way to involve kids is to let them have fun. **This is a key feature of this project; making it FUN and EASY to learn and develop e-learning.**

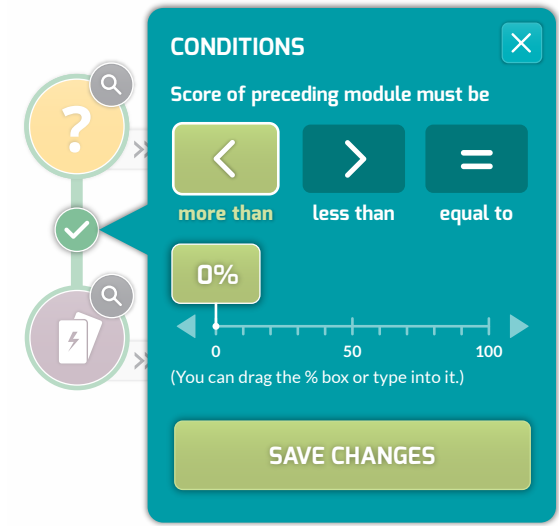

# THE SPECIFICS

## 01. Colours, fonts, and styles

**MAIN COLOURS**

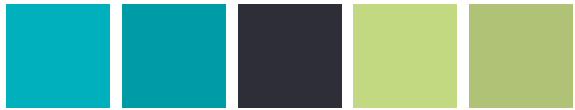

| MC1 | MC2 | MC3 | MC4 | MC5 |
|---|---|---|---|---|
| #00B0BC | #009CA7 | #2D2F37 | #C1DA83 | #AFC176 |

**SECONDARY COLOURS**

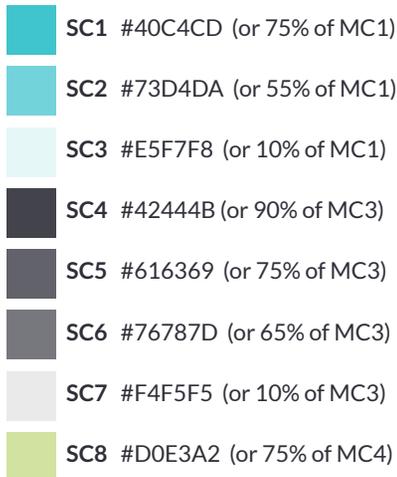

SC1  #40C4CD  (or 75% of MC1)

SC2  #73D4DA  (or 55% of MC1)

SC3  #E5F7F8  (or 10% of MC1)

SC4  #42444B  (or 90% of MC3)

SC5  #616369  (or 75% of MC3)

SC6  #76787D  (or 65% of MC3)

SC7  #F4F5F5  (or 10% of MC3)

SC8  #D0E3A2  (or 75% of MC4)

**FONTS**

### EXO
https://www.google.com/fonts/specimen/Exo

Used for most text

STYLES
Medium
Demi-bold
**Bold**

### Lato
https://www.google.com/fonts/specimen/Lato

Used for better readability of numbers, used in info-popups and tool-tips.

STYLES
Regular

**SHAPES AND SHADOWS**

The design is based on circles and rounded shapes. Shadows are used to indicate interactivity. Bigger shadows are used on most important buttons, darker shadows on dark backgrounds and lighter shadows on lighter backgrounds.

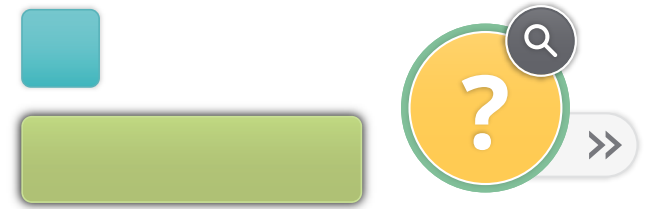

## 02. Initial view and interface

As already mentioned there are no splash-screens, loading or extra clicks to get into the system, you start playing right away.

To make it clear and simple, only a few actions are presented when you start off. The highlighted feature is the 'Find module' (**1**) as we want to encourage users to mainly re-use already existing modules. There is also a button to 'Make new' (**2**) module, which will use said modules authoring tool.

In addition you can 'OPEN' (**3**) a project, ask the avatar for 'HELP' (**4**), and view profile options by selecting the avatar picture. All unavailable functionality is clearly marked as such, and there is a textual hint that tells the user what they need to do to start.

If the system detects that a user spends a while without doing anything, it assumes the user is new, and a welcome message pops up. With a login-system or caching, the system would know only to do this the first time. By having a delay we avoid bothering users that actually know exactly what to do, and start off right away.

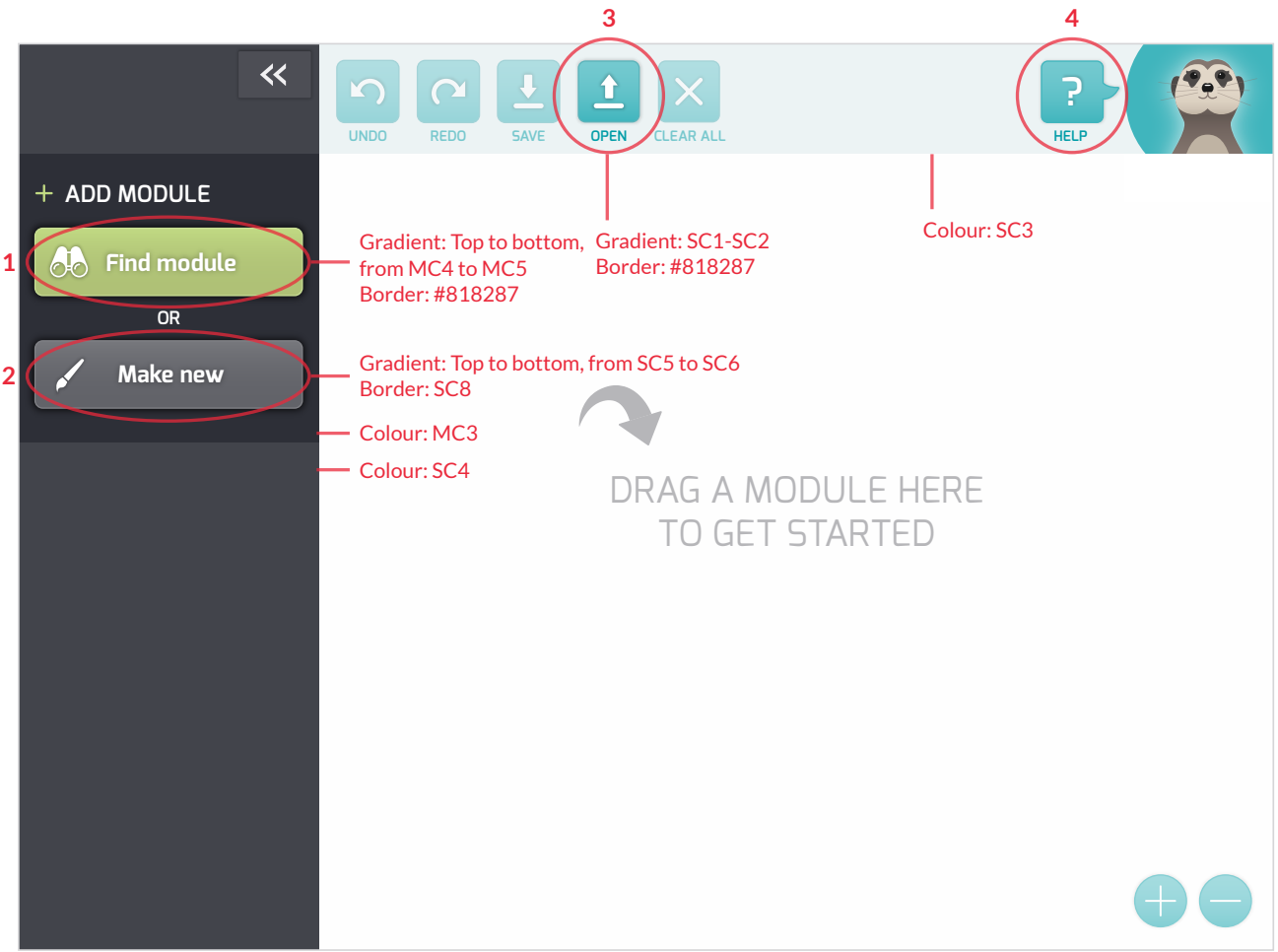

## 03. Additional functionality on initial view

Other functionality can be added to the initial view to motivate users, such as a list of some of their 'FAVOURITES' (**5**) or a list of their 'RECENTLY USED' (**6**) modules. Other functionality could include a list of the 'MOST POPULAR' modules, based on the influence score mentioned earlier.

Modules can be interacted with from these lists as well, indicated by the consistent design. If the user selects a module the module info box will appear (**7**), like it will other places in the interface. They can also drag these modules straight onto their canvas.

One problem that needs solving is how to suggest relevant modules. If a student is creating a course about Africa, then modules about algebra aren't very relevant, and serves more as a distraction than anything else. One possible solution is displaying *suggested modules* based on their search results.

Other functionality that becomes available when appropriate is 'UNDO' (**8**) — that reverses the last action, 'REDO' (**9**) — that undoes an undo, 'SAVE' (**10**) — that allows the user to save their project, and 'CLEAR ALL' (**11**) — which clears the whole canvas. In addition the canvas may be zoomed in and out, by either scrolling or by using the '+' and '−' buttons (**12**) in the lower right-hand corner.

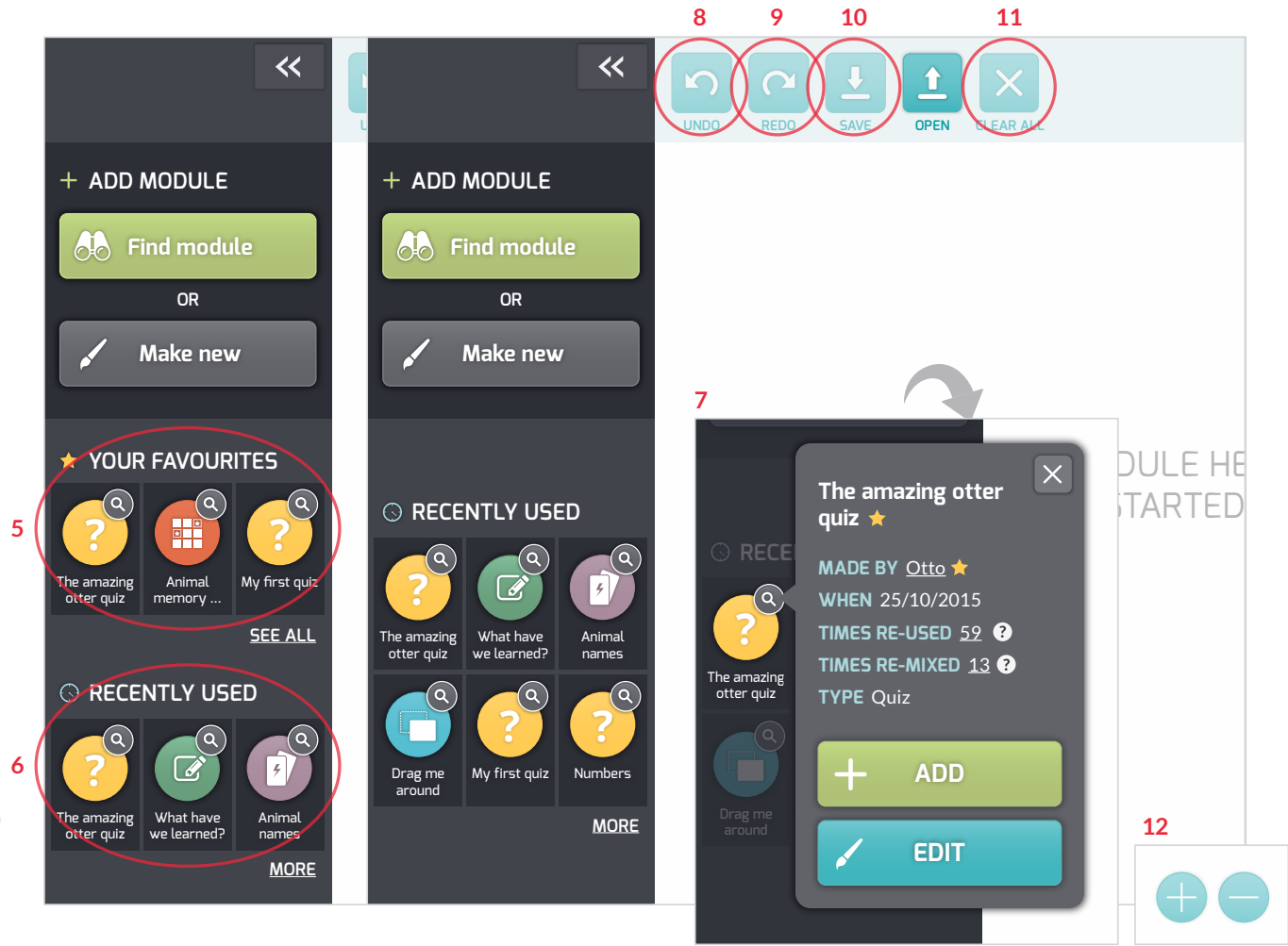

## 04. Expand/shrink the sidebar

To afford more room to the canvas, the left-hand side bar can be shrunk to a narrow bar with only icons. As a rule, we have text labels for all important functionality, but on the narrow bar we use only icons to save room. If the user is unsure of what any of the buttons do. they can simply expand the bar (13) to see the full buttons.

The 'FAVOURITES' (14), 'MOST POPULAR' (15), and 'RECENTLY USED' (16) buttons will result in the same action as the 'See all' (for 'FAVOURITES') or 'See more' (for 'MOST POPULAR' and 'RECENTLY USED').

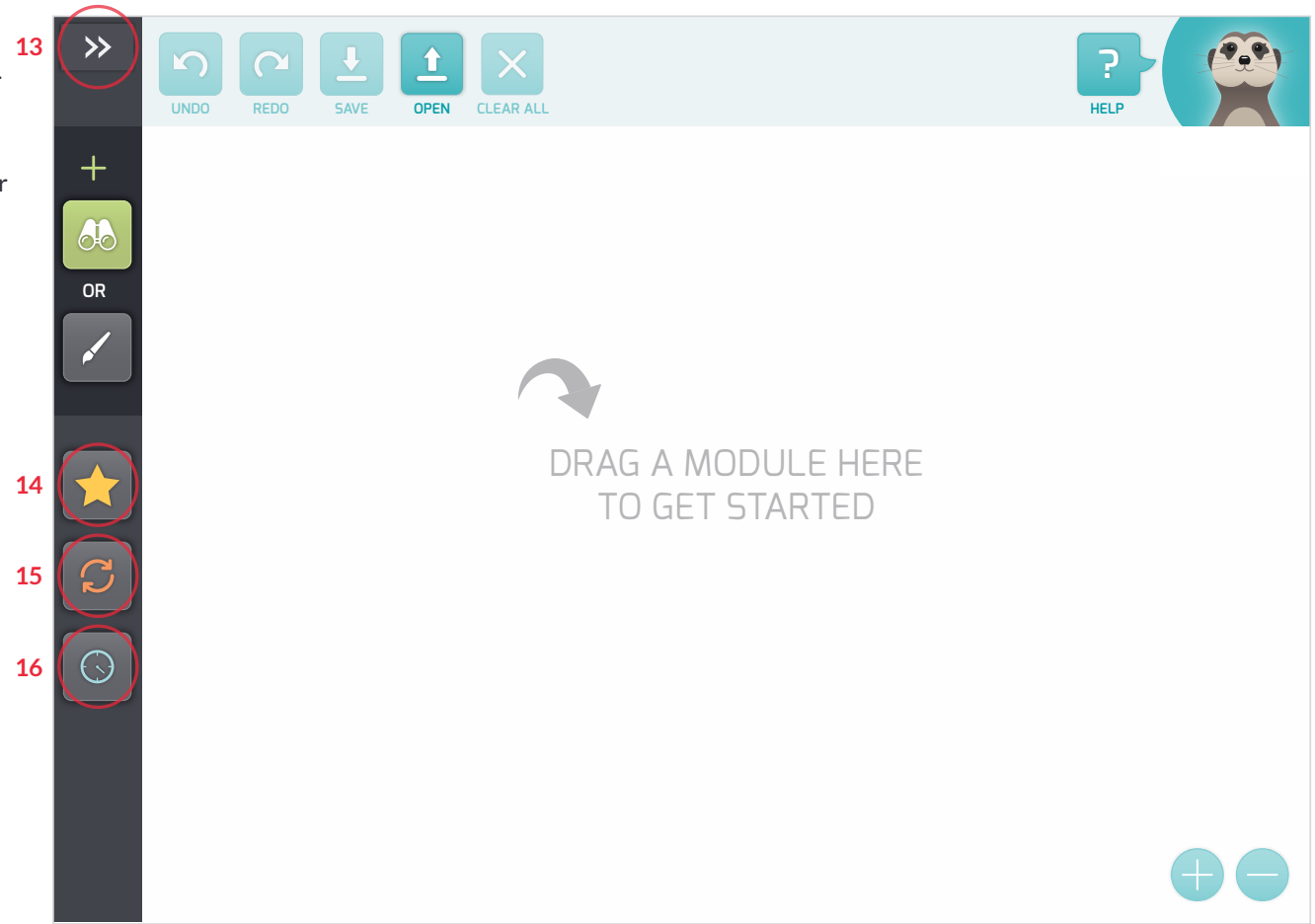

## 05. Find module

The 'Find module' panel is the only panel on the side bar that cannot be shrunk to a narrow version. Instead a 'CANCEL' (**17**) button appears. When selected the user is directed back to the previous screen. This panel takes up twice as much space as the normal side panel, to allow for extra information such as 'SEARCH FILTERS' (**18**) and 'RESULTS' (**19**).

The search field itself appears active and ready to be typed into when the user clicks the 'Find module' button, to avoid having them click a second time. (**20**)

The user can apply search filters before or after searching (**21 - next page**). Search results will be listed as module icons with the same appearance as elsewhere in the system (**22 - next page**), and can be dragged and dropped onto the canvas from there, or added by using the 'ADD' button (**23 - next page**).

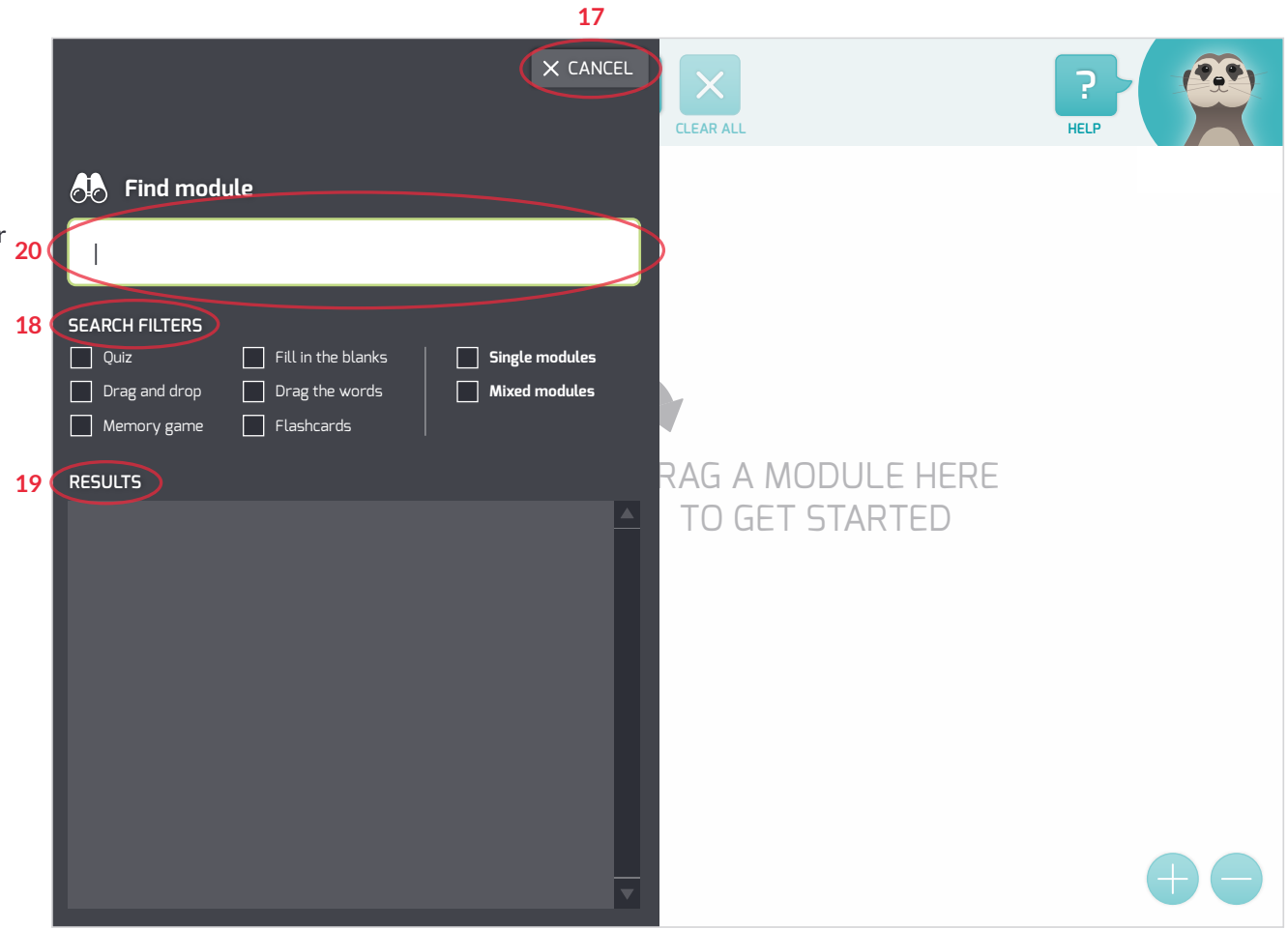

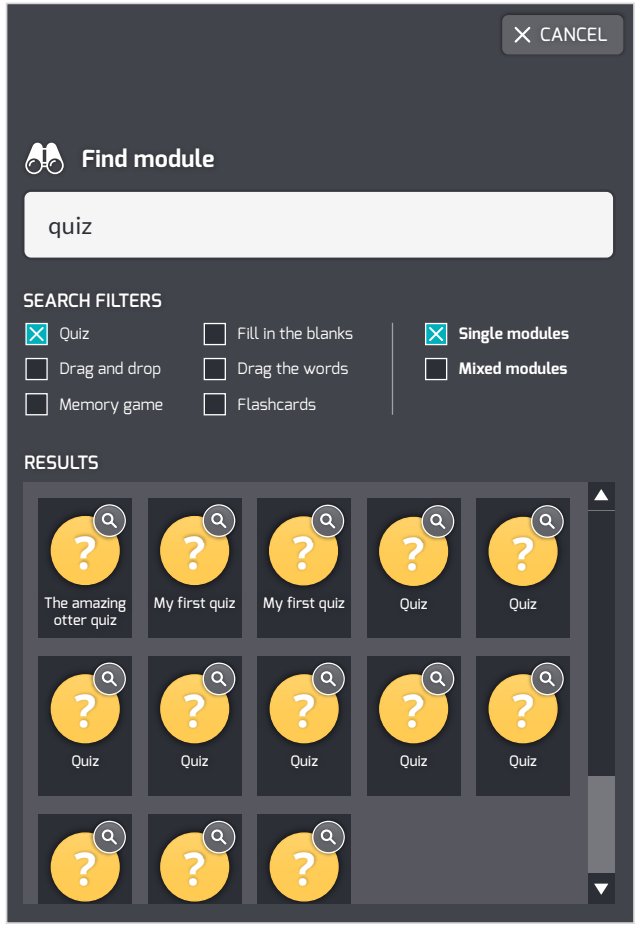
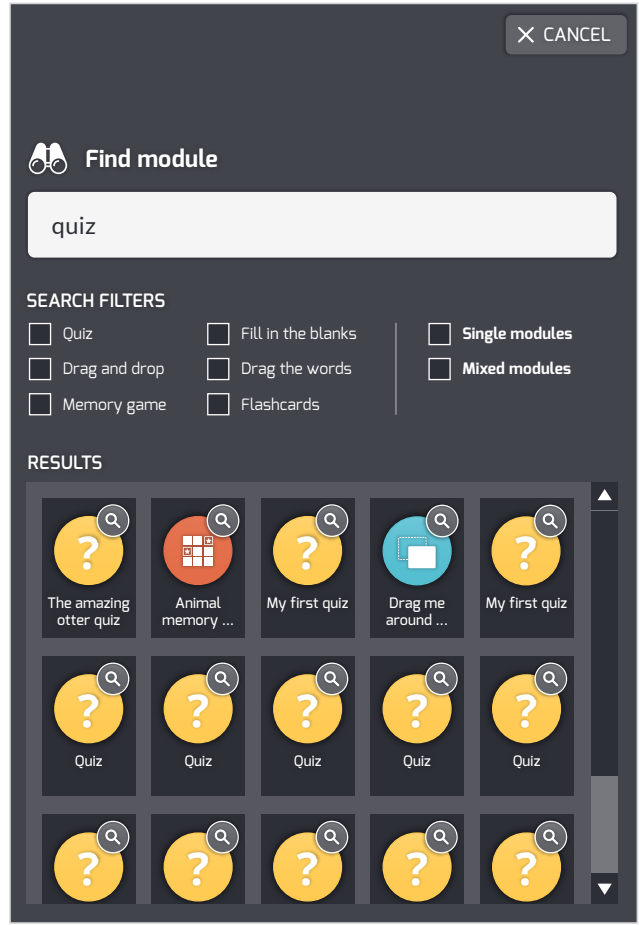
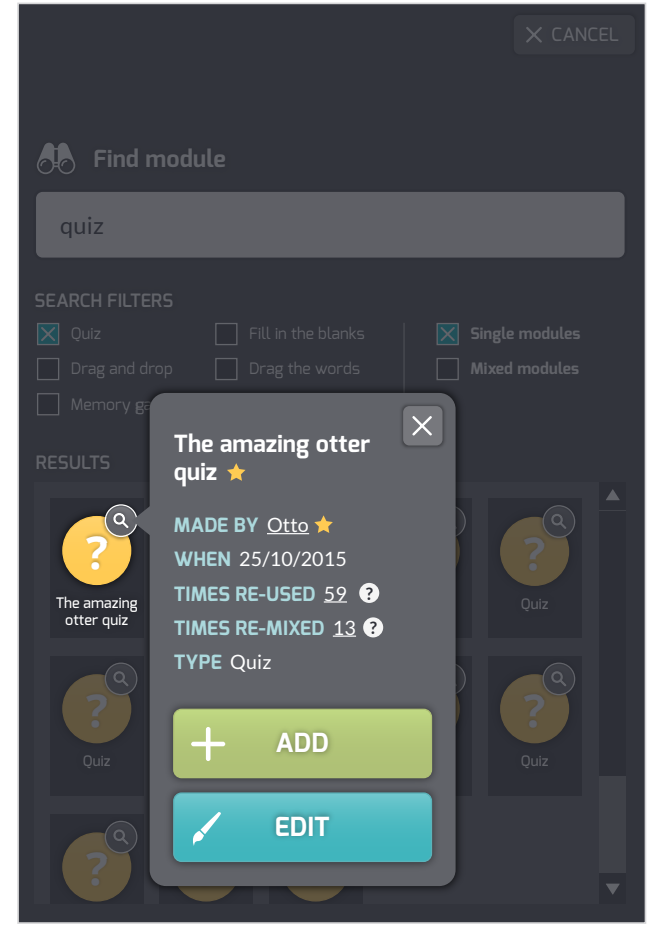

## 06. Make new module

The current interface for making new modules has 6 basic H5P-types. Quiz is the *question set* H5P-type, whilst the others retain their H5P default name and icon.

The 'Make'-panel can be shrunk into a smaller side-panel (**24**) with the same functionality.

Hitting 'CANCEL' (**25**) will take the user back to the previous screen. The 'Find module' button (**26**) is preserved in case the user changes their mind, and want to re-use an existing module instead of making a new one from scratch. It is less prominent and in a different location from the default view, but with the same look and size. This, along with morphing transitions, will indicate that the button is functionally equivalent in both places.

When the user chooses to add a module, they can either drag-and-drop the icon from the side bar onto the canvas (**27 - next page**), or they can select it and select the 'ADD' button that appears (**28 - next page**). The whole interface is darkened in response, so that the drop-zone is highlighted.

When dropped onto the canvas, the relevant module editor opens (**29 - next page**). Once the module is authored, it will appear on the canvas (**30 - next page**).

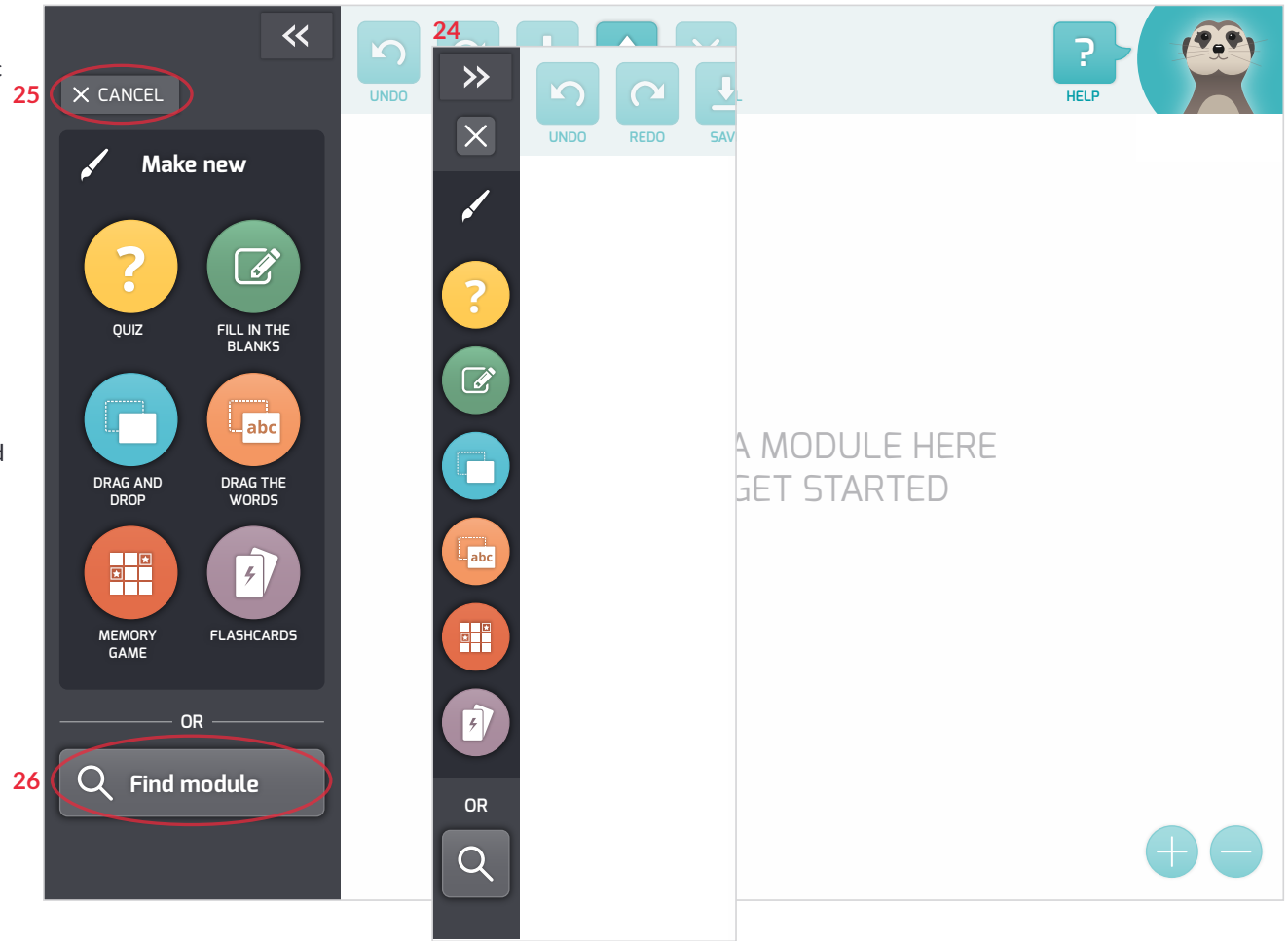

27

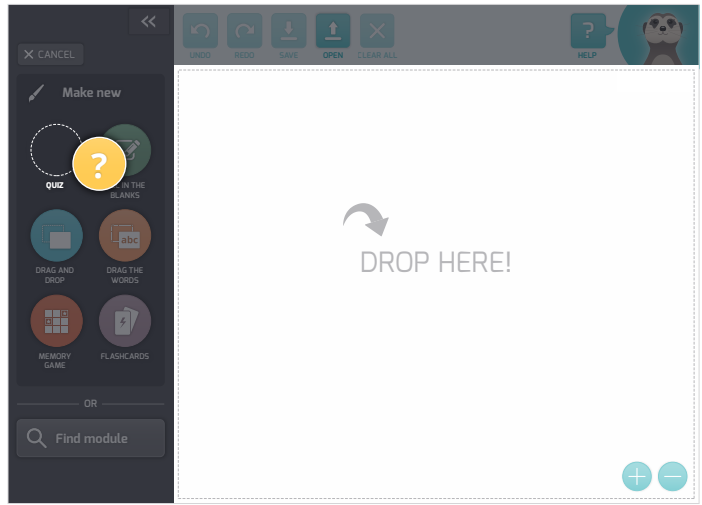

28

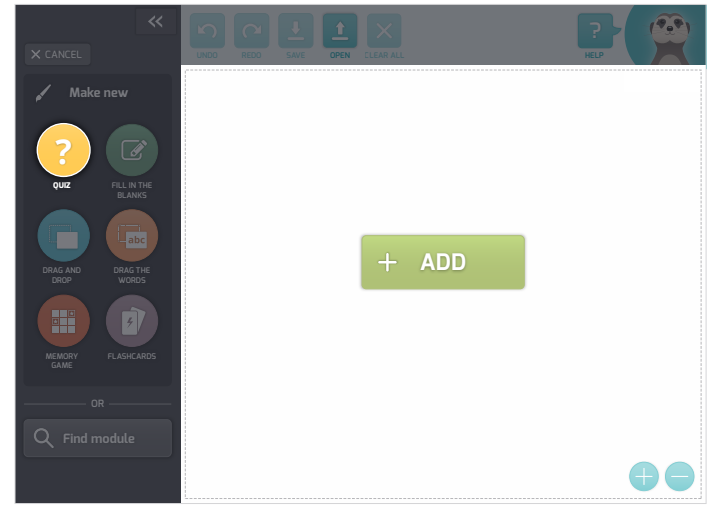

29

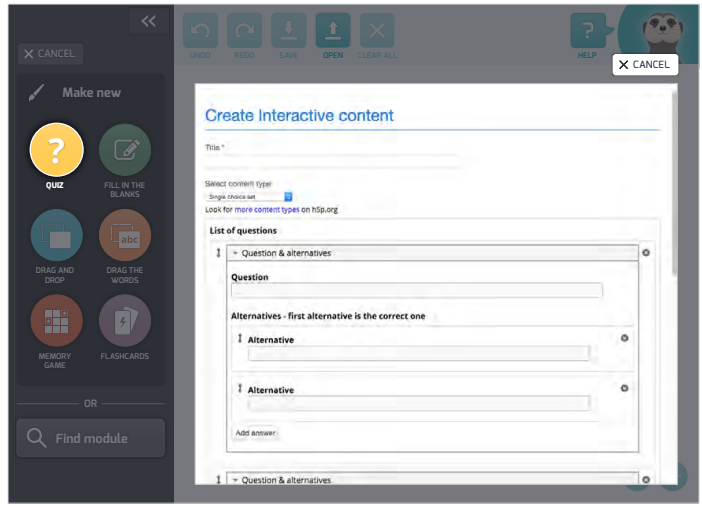

30

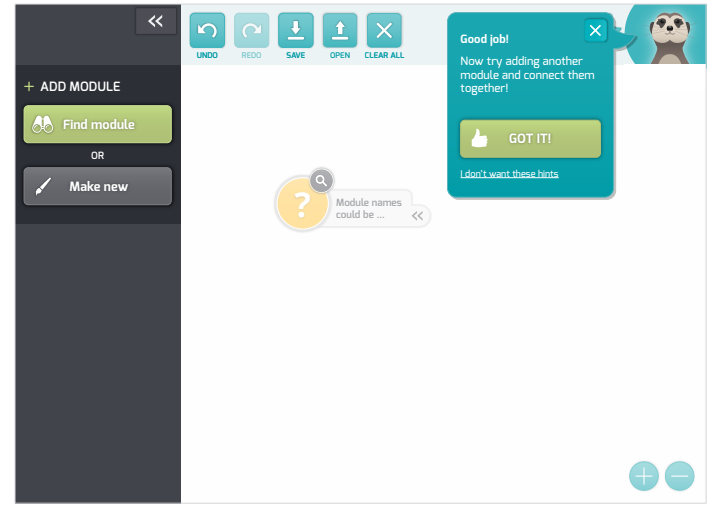

## 07. Make connections

When two or more modules are added to the canvas, the side-menu will change to include a new action: the 'CONNECTIONS' (**31**). This is a toggle-menu, where the cursor will change depending on which action is active. E.g. when the 'DRAW' functionality (**32**) is active, the cursor turns into a pencil (**33**) that allows the user to draw connections between modules.

Once the connection is successfully drawn (**34 - next page**), the system will align the connected modules appropriately to create a more neatly organised system. (**35 - next page**). With only one module connected to the preceding module, the default condition on this connection will be that the second module follows the first, or that the score of the preceding module needs to be more than 0% (**36 - next page**). This can be overruled by the user, to allow for complete freedom, but the system will then indicate that the logic is broken for that particular connection, and that they need to fix it. The avatar will show up with a friendly warning the first time this happens (**37 - next page**), as well as the visual indicators on the module structure itself, with the connection turning red, and a triangle exclamation mark appearing in place of the green tick icon.

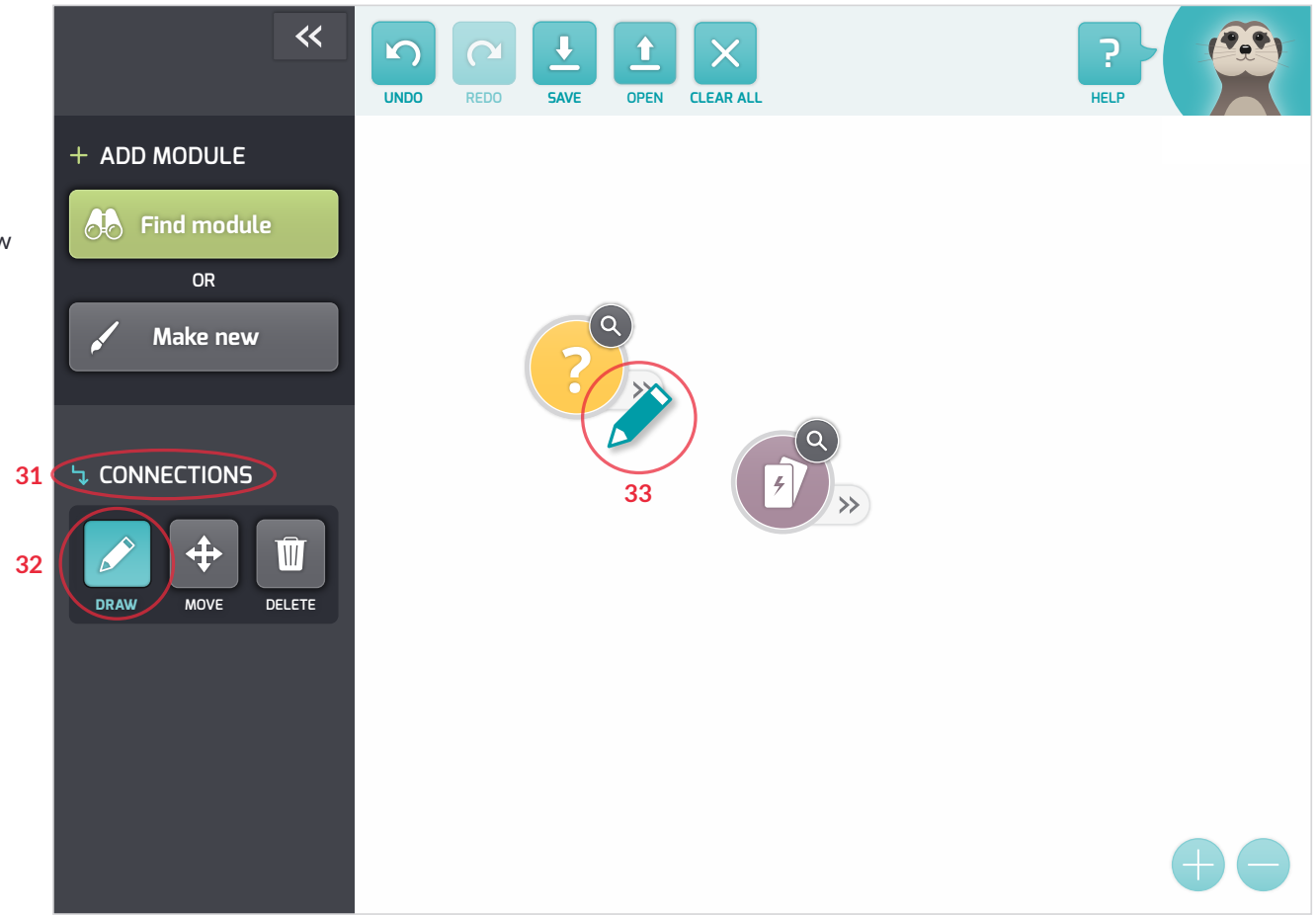

34

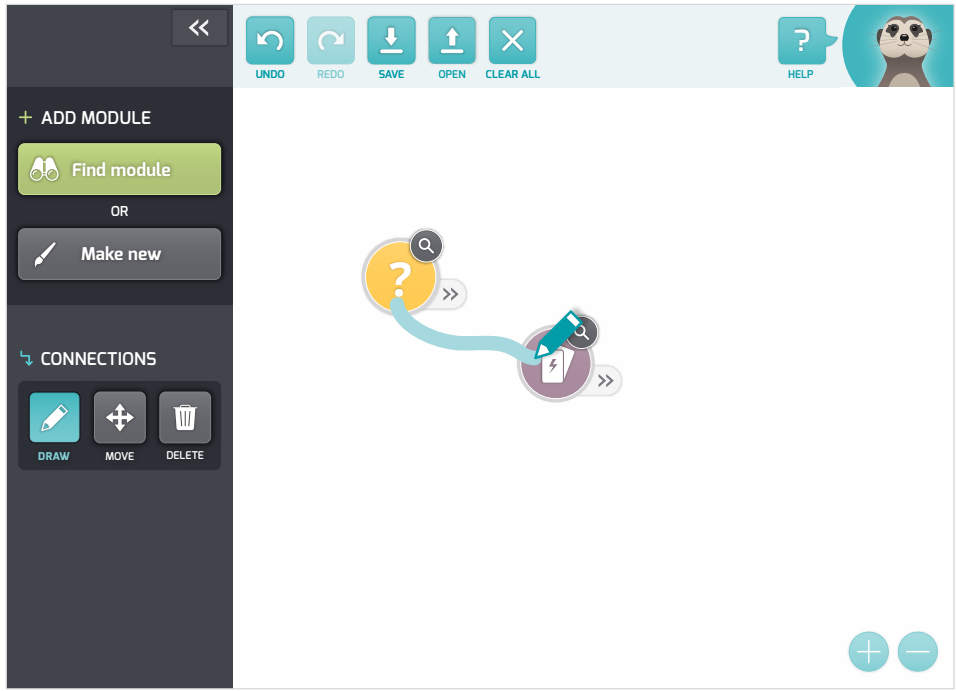

35

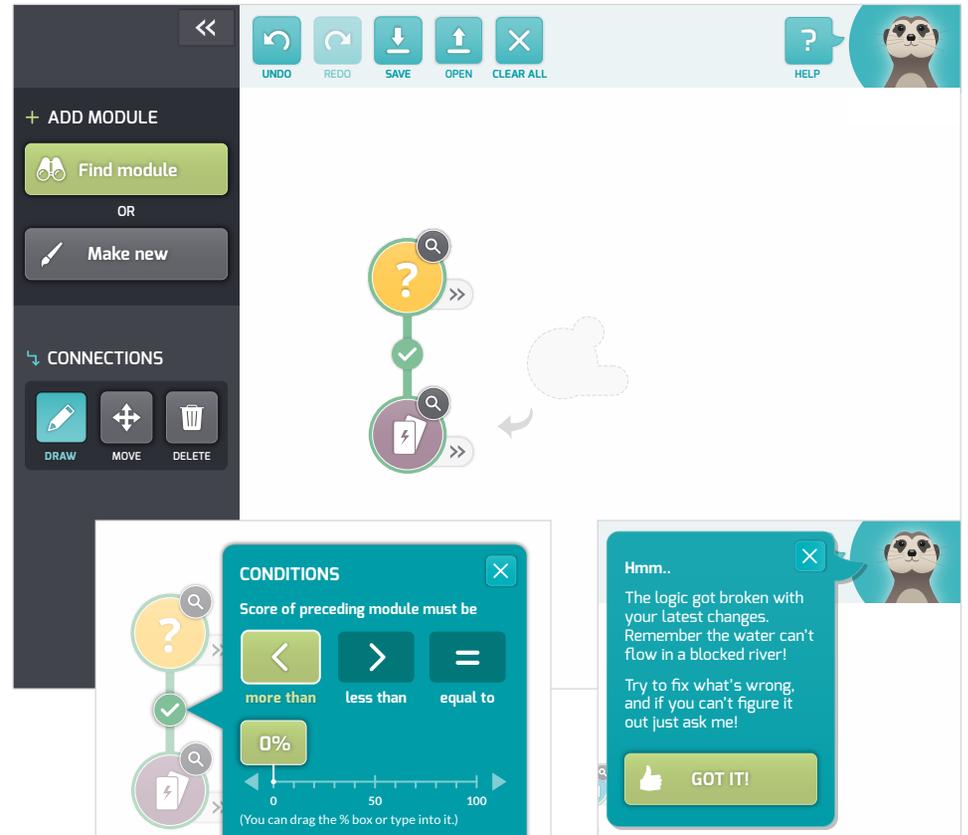

36

37

## 08. Demo

When a module is selected (**38**), the module info-panel will appear. This allows the user to see details about the current module, such as who it was made by, when it was made, how many times it has been re-used and re-mixed, and what type of module it is. From here, they may also edit or delete the module.

Additionally, when a module that has connections going from it is selected, a new menu item called 'CONDITIONS' (**39**) appears. This describes how many conditioned connections this specific module has. A button that allows the user to view these conditions (**40**) also appear. This will highlight the conditions that belongs to this module.

When a lot of modules are added the canvas may look cluttered on a small screen (**41 - next page**). One way to see more of the canvas, is by shrinking the side panel, (**42 - next page**) or by zooming out (**43 - next page**) the canvas. The heads-up display notifications retains their dimensions in order to preserve readability.

If a user's device has a large screen, the interface will adapt by giving them a larger area to operate in (**44 - next page**).

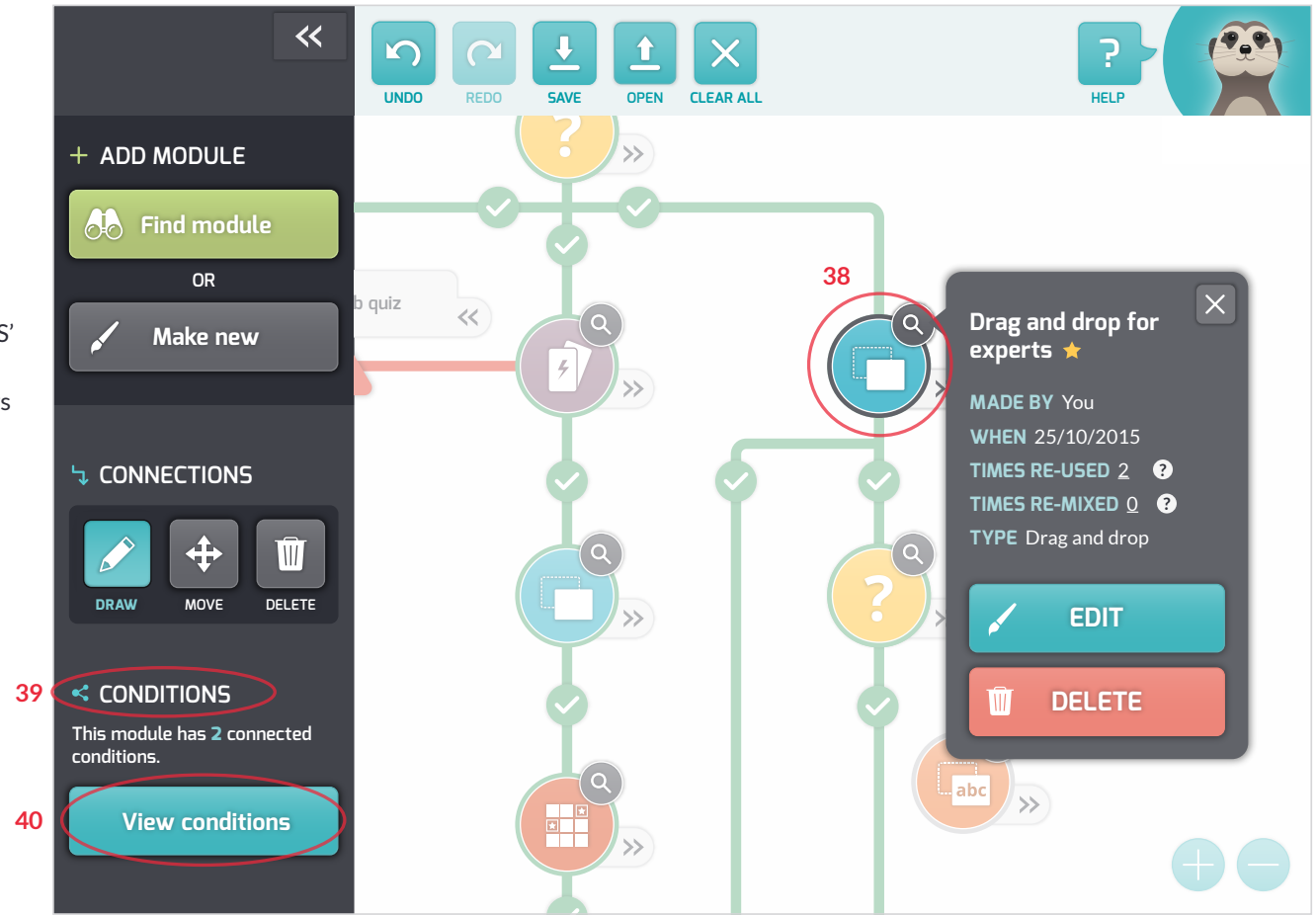

41 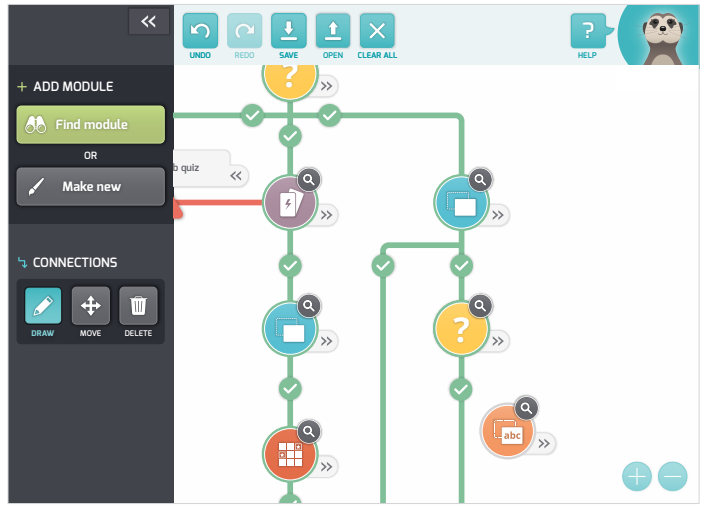

42 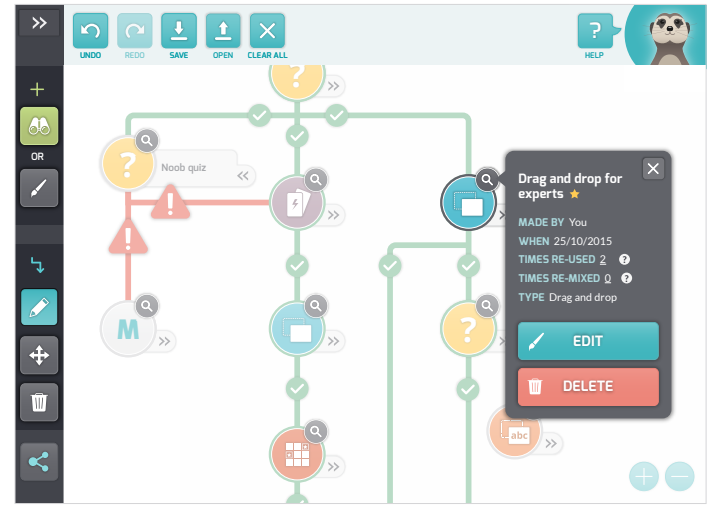

43 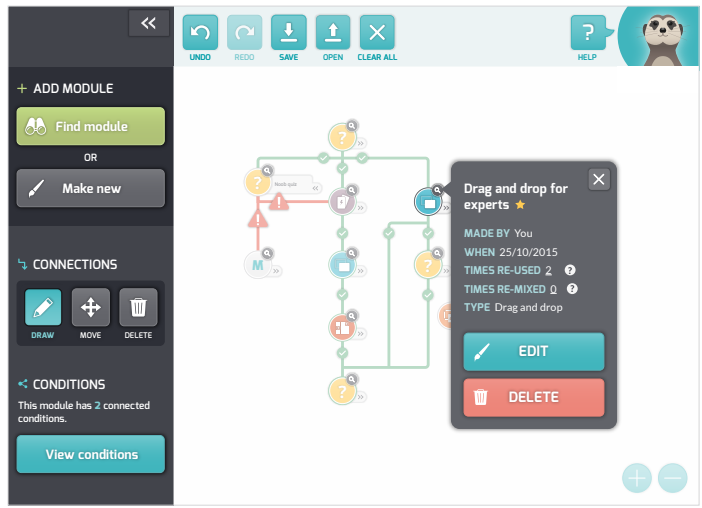

44 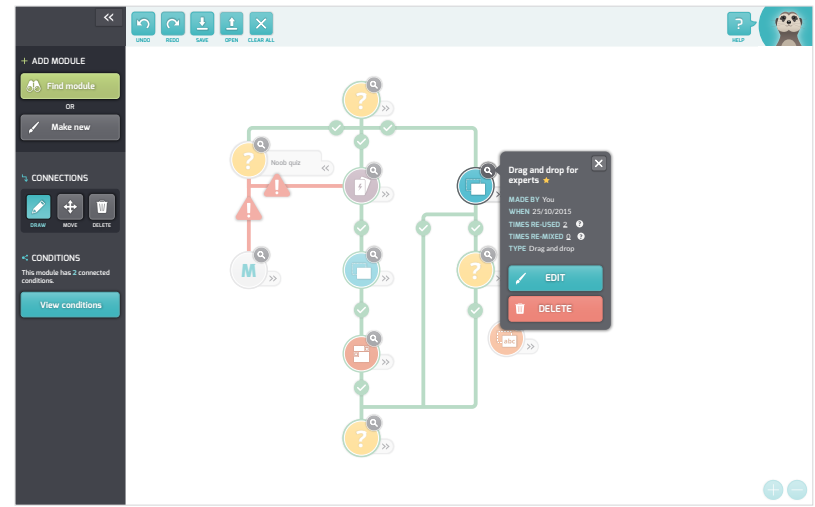

# REFERENCES

1. James Ignacio. Linear vs. Non-linear. https://ionelearning.wordpress.com/2013/07/12/linear-vs-non-linear/, 2013. Visited 2015-10-09.

2. Alexander Berntsen, Stian Ellingsen & Emil H Flakk. Enabling Learning By Teaching: Intuitive Composing of E-Learning Modules. 2015.

3. Jerry Cao, Kamil Zieba & Matt Ellis. UXPin: Interaction Design Best Practices. Mastering Words, Visuals, Space. E-book


\end{document}